\documentclass[preprint,journal]{vgtc}            

\onlineid{0}

\vgtccategory{Research}

\vgtcpapertype{system}

\title{ProvenanceWidgets: A Library of UI Control Elements\\ to Track and Dynamically Overlay Analytic Provenance}

\author{%
  \authororcid{Arpit Narechania}{0000-0001-6980-3686},
  \authororcid{Kaustubh Odak}{0009-0006-5873-7891},
  \authororcid{Mennatallah El-Assady}{0000-0001-8526-2613}, and
  \authororcid{Alex Endert}{0000-0002-6914-610X}
}

\authorfooter{
  \item
  	Arpit Narechania, Kaustubh Odak, and Alex Endert are with Georgia Tech. 
  	E-mails: \{anarechania3, kodak, endert\}@gatech.edu.
  \item
  	Mennatallah El-Assady is with ETH Zürich.
  	E-mail: melassady@ai.ethz.ch.
	}

\abstract{%
We present ProvenanceWidgets, a Javascript library of UI control elements such as radio buttons, checkboxes, and dropdowns to track and dynamically overlay a user's analytic provenance. These in situ overlays not only save screen space but also minimize the amount of time and effort needed to access the same information from elsewhere in the UI. In this paper, we discuss how we design modular UI control elements to track how often and how recently a user interacts with them and design visual overlays showing an aggregated summary as well as a detailed temporal history. We demonstrate the capability of ProvenanceWidgets by recreating three prior widget libraries: (1) Scented Widgets, (2) Phosphor objects, and (3) Dynamic Query Widgets. We also evaluated its expressiveness and conducted case studies with visualization developers to evaluate its effectiveness. We find that ProvenanceWidgets enables developers to implement custom provenance-tracking applications effectively. ProvenanceWidgets is available as open-source software at \url{https://github.com/ProvenanceWidgets} to help application developers build custom provenance-based systems.
}


\teaser{
  \centering
  \includegraphics[width=\linewidth, alt={ProvenanceWidgets}]{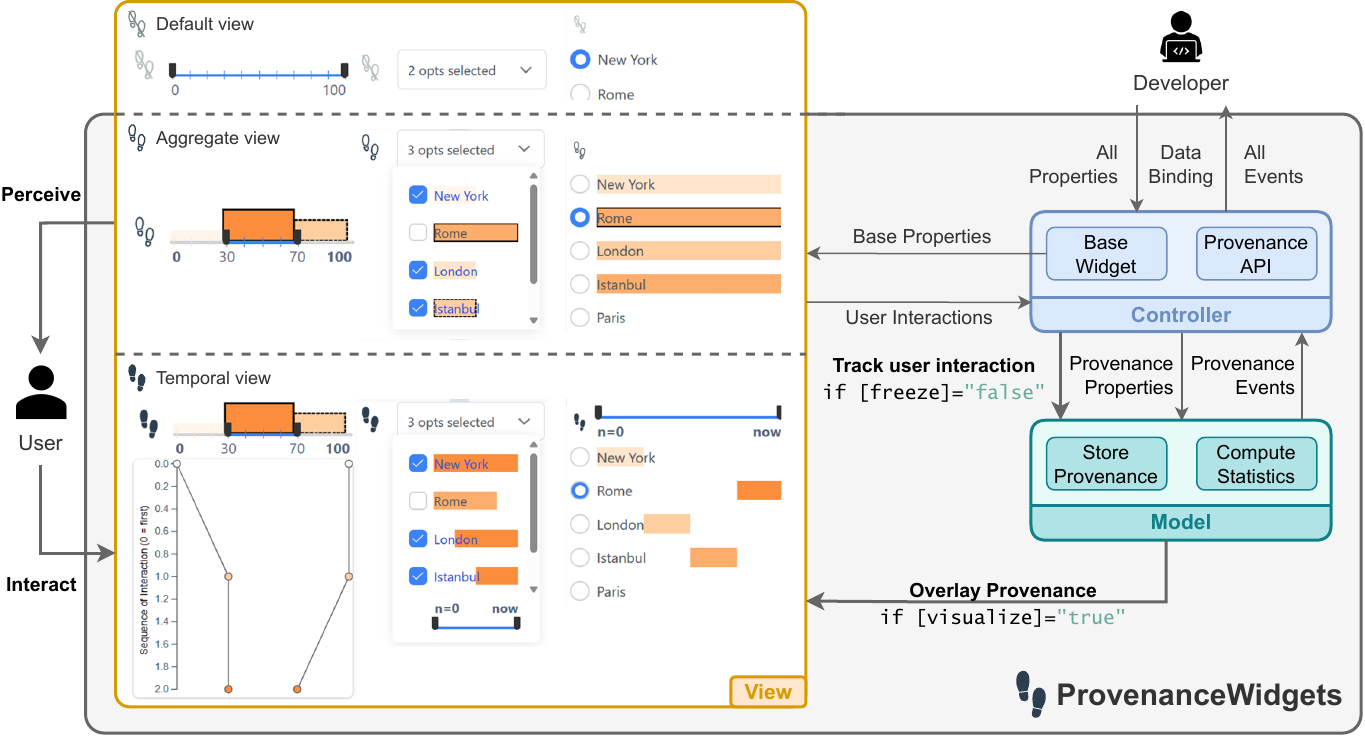}
  \caption{Overview of \app and the underlying \model{}-\view{}-\controller{}-based architecture. The \model{} stores, computes, and updates the provenance. The \view{} shows how end-users perceive and interact with the widgets. The \controller{} describes how the \model{}, \view{}, and developers can interact with \app.}
  \label{fig:teaser}
}

\graphicspath{{figs/}{figures/}{pictures/}{images/}{./}} 

\usepackage{tabu}                      
\usepackage{booktabs}                  
\usepackage{lipsum}                    
\usepackage{mwe}                       

\usepackage{mathptmx}                  

\usepackage{enumitem}
\usepackage{soul}
\usepackage{listings}
\usepackage{quoting}

\newcommand{\bpstart}[1]{
\noindent{\textbf{#1}}%
}

\newcommand{\app}{ProvenanceWidgets\xspace}

\newcommand{\paragraphHeadingSpace}{\vspace{4px}}

\usepackage{setspace} 
\lstdefinelanguage{TypeScript}{
    basicstyle=\small\ttfamily\color{blue}
}

\lstdefinelanguage{HTML5}{
    language=HTML,
    basicstyle=\small\ttfamily\color{magenta}
}

\lstdefinelanguage{CSS}{
    basicstyle=\small\ttfamily\color{orange}
}

\lstdefinelanguage{Angular}{
    language=HTML,
    basicstyle=\small\ttfamily\color{magenta},
    alsoletter={<>=-},
    morekeywords={
        provenance-,
        slider,
        dropdown,
        multiselect,
        radiobutton,
        checkbox,
        inputtext,
        provenance-slider,
        provenance-inputtext,
        provenance-dropdown,
        provenance-multiselect,
        provenance-radiobutton,
        provenance-checkbox
    },
    deletekeywords={value, selected, data}
}

\lstdefinelanguage{TS}{
    language=Java,
    morekeywords={
        SliderProvenance,
        InputTextProvenance,
        Provenance,
        "interaction",
        "time",
        number,
        string,
        ChangeContext,
        Options,
        Option,
    },
    deletekeywords={label, true, if, return}
}

\lstdefinestyle{TypeScript}{
  language=TypeScript,
  basicstyle=\small\ttfamily,
  keywordstyle=\color{blue},
  commentstyle=\color{green!60!black},
  stringstyle=\color{red},
  breaklines=true,
  showstringspaces=false,
  numbers=left,
  numberstyle=\tiny,
  frame=lines,
  captionpos=b
}

\lstdefinestyle{HTML}{
  language=HTML5,
  basicstyle=\small\ttfamily,
  keywordstyle=\color{blue},
  commentstyle=\color{green!60!black},
  stringstyle=\color{red},
  breaklines=true,
  showstringspaces=false,
  numbers=left,
  numberstyle=\tiny,
  frame=lines,
  captionpos=b
}

\lstdefinestyle{Angular}{
  language=Angular,
  basicstyle=\small\ttfamily,
  keywordstyle=\color{blue},
  commentstyle=\color{darkgray},
  stringstyle=\color{teal},
  breaklines=true,
  showstringspaces=false,
  numbers=left,
  numberstyle=\tiny,
  frame=topline,
  captionpos=b,
}

\lstdefinestyle{TS}{
  language=TS,
  basicstyle=\small\ttfamily,
  keywordstyle=\color{teal},
  commentstyle=\color{lightgray},
  stringstyle=\color{teal},
  breaklines=true,
  showstringspaces=false,
  numbers=left,
  numberstyle=\tiny,
  frame=none,
  belowskip=0em,
  captionpos=b
}

\newcommand{\checkboxicon}[1]{\includegraphics[width=2em]{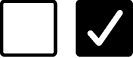}}

\newcommand{\radiobuttonicon}[1]{\includegraphics[width=2em]{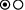}}

\newcommand{\slidericon}[1]{\includegraphics[width=3em]{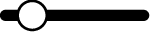}}

\newcommand{\rangeslidericon}[1]{\includegraphics[width=3em]{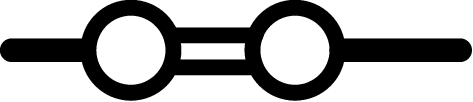}}

\newcommand{\singleselecticon}[1]{\includegraphics[width=3em]{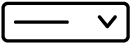}}

\newcommand{\multiselecticon}[1]{\includegraphics[width=3em]{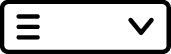}}

\newcommand{\inputtexticon}[1]{\includegraphics[width=3em]{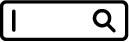}}

\newcommand{\aggmodeicon}[1]{\includegraphics[width=10pt]{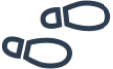}}
\newcommand{\tempmodeicon}[1]{\includegraphics[width=10pt]{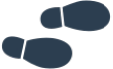}}
\newcommand{\disabledmodeicon}[1]{\includegraphics[width=10pt]{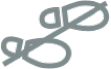}}

\newcommand{\user}[1]{\textcolor{teal}{$P_{#1}$}}

\newcommand{\model}[0]{\textcolor{teal}{Model}}
\newcommand{\view}[0]{\textcolor{orange}{View}}
\newcommand{\controller}[0]{\textcolor{blue}{Controller}}

\newcommand{\cut}[1]{}
\newcommand{\edit}[1]{#1}
\newcommand{\add}[1]{#1}

\begin{document}


\firstsection{Introduction}

\maketitle

\label{section:introduction}
Analytic provenance is the documented history of data and analytical actions, showing how data was obtained, transformed, and analyzed. 
In a visualization context, analytic provenance tracks how users interact with visualizations as a representation of their reasoning process~\cite{north2011analytic} and can be helpful for recalling the analysis process, reproducing it, collaborating, and logging for evaluation or meta-analysis~\cite{ragan2015characterizing}.
Presenting provenance during analysis has been shown to increase awareness of analytic behaviors~\cite{narechania2021lumos,paden2024biasbuzz}, increase confidence~\cite{block2022influence}, reduce exploration biases~\cite{wall2021lrg}, and result in more unique insights~\cite{feng2016hindsight,willett2007scented}, among others.

Prior work has made strides in libraries that help developers capture and store provenance~\cite{callahan2006vistrails,aigner2013evalbench, okoe2015graphunit,cutler2020trrack}.
For example, Trrack~\cite{cutler2020trrack} is an open-source library to create and track the provenance (history) of interactions in web-based applications for a variety of purposes such as action recovery, reproducibility, collaboration, and logging.

While logging and analyzing provenance after analysis has immense value, there is a need for libraries that aid developers in integrating provenance into visual analytic tools in a manner that is consistent with common UI standards.
There exist many open-source libraries of UI controls~\cite{materialui, reactbootstrap, vuetify, angularmaterial, primeng, sveltekitui, jqueryui, bootstrap} that enhance user interaction and facilitate data input in software applications or websites.
Even in visualization and HCI research, there are libraries of enhanced UI controls that facilitate data exploration~\cite{williamson1992dynamic,heer2008generalized,vaithilingam2024dynavis} and navigation~\cite{willett2007scented}, explain transitions in the user interface via afterglow effects~\cite{baudisch2006phosphor}, or visualize group awareness information~\cite{hill2003awareness}.
However, there is no frontend library of UI controls that tracks and presents provenance information \emph{out of the box} during analysis.
Tools like TrrackVis (Trrack's~\cite{cutler2020trrack} frontend library) visualize the logged provenance information graph; however, these visualizations are available in a separate view/tool, sometimes after analysis.
So we ask: \textit{how can developers integrate provenance directly into the user interface of visual data analysis tools?}

In response, we present \textbf{\app}, a Javascript library of UI control elements such as radio buttons and dropdowns to track and dynamically overlay a user's analytic provenance, \emph{out of the box}. 
These enhanced widgets track how often (frequency) and when (recency) a user interacts with them \edit{(e.g., selecting a dropdown option)} and present visual overlays showing an aggregated summary as well as a detailed temporal history.
The aggregated summary is presented in a bar chart overlay visualization encoding the frequency (length) and recency (color) of user interactions with the widget.
The detailed temporal history is presented as a timeline visualization, enabling users to access granular information about specific interactions in the past.
Presenting provenance as overlays not only saves screen space but also minimizes the amount of time and effort needed to access the same information from other views or external tools.

The ProvenanceWidgets library consists of radio buttons~\radiobuttonicon{}, checkboxes~\checkboxicon{}, single sliders~\slidericon{}, range sliders~\rangeslidericon{}, dropdowns~\singleselecticon{}, multiselects~\multiselecticon{}, and input text fields~\inputtexticon{}.
The library is built using Angular but is universally compatible across different frameworks through Web Components.
The library is also highly customizable, allowing developers to realize a variety of configurations such as setting the logging frequency (e.g., logging every interaction or every second), or initializing with a provenance log previously captured.

We demonstrate the capability of ProvenanceWidgets through two usage scenarios and recreate three prior widget libraries: (1) Scented Widgets, (2) Phosphor objects, and (3) Dynamic Query Widgets. 
In addition, we evaluate the expressiveness of \app through cognitive dimensions of notations.
We also evaluate the effectiveness of \app through case studies with four visualization developers. 
We find that \app enables developers to effectively implement custom provenance-tracking applications. 
ProvenanceWidgets is available as open-source software at \url{https://github.com/ProvenanceWidgets} to help visualization developers create new or enhance existing provenance-based systems.

\section{Related Work}
\label{section:related-work}
\subsection{Analytic Provenance}
Our memory has a limited capacity to remember and track our prior interactions with data, in both amount and decay~\cite{miller1994magical, liu2014effects}, which creates a barrier to data exploration.
Analyzing prior interactions with data in visualization is a form of analytic provenance~\cite{ragan2015characterizing,north2011analytic} that is often used to infer one's analysis process.
In fact, provenance has been found to play a crucial role in supporting decision-making, fostering collaboration, and enhancing understanding of complex data analysis in visualization and data analysis contexts~\cite{madanagopal2019analytic}.
Provenance has also been shown to affect users' confidence levels in conclusions developed, the propensity to repeat work, filtering of data, identification of relevant information, and during typical investigation strategies~\cite{block2022influence}.

\paragraphHeadingSpace\bpstart{Tracking Analytic Provenance.} 
Provenance can be tracked in workflow modeling systems~\cite{bavoil2005vistrails} as well as the analysis process within an interactive system~\cite{north2011analytic}.
There have been many provenance-logging frameworks~\cite{callahan2006vistrails,aigner2013evalbench,okoe2015graphunit,cutler2020trrack} that have been used to improve empirical evaluations of visualization techniques~\cite{revisit2023ding,nobre2021revisit} or teach scientific visualization~\cite{silva2011using}, among others.
VisTrails~\cite{callahan2006vistrails} is an open-source system that manages the data and metadata of visualization products by capturing the provenance of both the visualization processes and the data they manipulate.
Trrack~\cite{cutler2020trrack} is an open-source library to capture and replay complete provenance graph; Trrack is complemented by a frontend library (TrrackVis) for visualizing and interacting with this data.

\paragraphHeadingSpace\bpstart{Visualizing Analytic Provenance.}
Visualizing one's prior interactions can take many forms and lead to shifts in a user's analysis behavior.
For instance, when users' prior interactions with charts or data points are encoded (e.g., analogous to coloring previously visited hyperlinks on a webpage purple), people tend to interact with more unique (or previously unvisited) data~\cite{feng2016hindsight} or the same data repeatedly~\cite{narechania2021lumos,wall2021lrg}.
Similarly, when exploration history is shown in interactive network visualizations, users report inspiration for conducting further analysis and greater recall of their prior explorations~\cite{dunne2012graphtrail}.
Scientists have also been found to efficiently and effectively explore their visualizations by returning to previous versions of a dataflow (or visualization pipeline)~\cite{callahan2006vistrails}.
Recently, computational notebooks have leveraged provenance information to provide direct feedback on the impact of changes made within cells for iterative and exploratory data analysis~\cite{2024_loops, epperson2022leveraging}.
Well-designed histories can also help users maintain contextual awareness of previously visited data when distortions are applied that would otherwise make contextual awareness a challenge (e.g., fisheye lens)~\cite{skopik2005improving}. 

Graphical traces of user interactions have also been utilized in collaborative visualization settings, e.g., to facilitate coordination of multiple users by showing current selections and interactions as ``coverage'' of the data~\cite{sarvghad2015exploiting,badam2017supporting} and in personalized integrated development environments (IDEs), e.g., Footsteps for VSCode~\cite{footstepsvscode} highlights the lines of code as the user edits them.
Similarly, showing social information ``scents'' on data visualization widgets (e.g., representing others' interactions with radio buttons, sliders, etc.) leads users to make substantially more unique discoveries in the data~\cite{willett2007scented}.

Traces of prior interactions have also been applied in HCI contexts, including tracking user focus while browsing a webpage using eye-tracking\cite{nielsen2010eyetracking} and mouse-tracking\cite{arroyo2006usability} gear, tracking interactions with documents by authors and readers~\cite{hill1992edit}, facilitating groupware coordination~\cite{gutwin2002traces}, revisiting common regions of a page using scrollbar history~\cite{alexander2009revisiting}, among others.
We refer to Heer~et~al.~\cite{heer2008graphical} for a detailed review of the design space for displaying interaction histories.

\subsection{\add{Web and Gaming Analytics}}

\add{Recently, video game analytics has gained popularity, utilizing game telemetry data~\cite{kohwalter2017capturing, drachen2015behavioral} to measure player performance~\cite{chung2015guidelines}, 
characterize player behavior~\cite{drachen2013game, melo2020player, gagne2011deeper} and experiences~\cite{drachen2015behavioral, jacob2017oh}, 
understand and improve game design~\cite{drachen2015behavioral, kohwalter2018understanding, kohwalter2020provchastic, jacob2017oh},
and explore social phenomenon~\cite{lim2015toward}. 
For example, Melo~et~al.~\cite{melo2020player} proposed profiling player behavior through provenance graphs and representation learning.
Provchastic~\cite{kohwalter2020provchastic} analyzed game dynamics to predict future game events. 
Jacob~et~al.~\cite{jacob2017oh} utilized sequential pattern mining algorithms to identify cycles in a game pattern, making the game less tedious and also informing difficulty adjustments. 
AIRvatar~\cite{lim2015toward} studied how click events and time durations from users' customization of avatars can highlight gender-related stereotypes~\cite{lim2015toward}.}

\add{Web-based analytics~\cite{googleanalytics,mixpanel, newrelic} and session replay~\cite{hotjar, fullstory, mouseflow} tools also aid tracking and visualization of user interactions and behavior on websites.
These tools capture a variety of metrics such as web page load times, network response times, error rates, including user interactions such as hovers, clicks, and scrolls with the page elements.
Beyond low-level data collection, these tools also synthesize high-level user patterns and behaviors that aid performance monitoring (e.g., detect trends and anomalies) and enhance user engagement (e.g., provide guidance). 
These tools also employ advanced visualizations such as heatmaps to highlight areas of user interaction on web pages, helping optimize their layout and content placement (i.e., facilitate A/B testing)~\cite{hotjar, mouseflow}.}

\add{In contrast, \app tracks and visualizes the provenance of user interactions that alter the value of UI controls, independent of broader context like web page load times or game events.}

\subsection{UI Controls and Libraries}

There are many open-source libraries of UI controls that enhance user interaction and facilitate data input in software applications or websites~\cite{materialui, reactbootstrap, vuetify, angularmaterial, primeng, sveltekitui, jqueryui, bootstrap}.
By utilizing these libraries, developers can expedite their development process while ensuring consistency and accessibility across various platforms and devices.

Visualization and HCI researchers have also developed several enhanced UI control libraries.
For example, Phosphor objects~\cite{baudisch2006phosphor} instantly show and explain state transitions in GUI controls, e.g., manipulating a phosphor slider leaves an afterglow that illustrates how the knob moved.
Scented Widgets~\cite{willett2007scented} enhance GUI controls with embedded visualizations that facilitate navigation in information spaces.
\add{Emotion scents~\cite{cernea2013emotion} tracks users' emotional reactions while interacting with GUI widgets and visualizes these reactions on the widgets, enhancing the interface for emotional awareness and decision support.}
TrrackVis provides a customizable provenance visualization front-end for the Trrack library~\cite{cutler2020trrack}.
DynaVis~\cite{vaithilingam2024dynavis} synthesizes persistent UI widgets in response to an initial natural language-based visualization editing task, enabling the user to make subsequent modifications by directly interacting with the widgets (instead of re-typing natural language).

\add{In contrast, \app is a library of UI controls with built-in provenance tracking and visualization overlays, and APIs to integrate the provenance information into broader application workflows. }

\section{Provenance Widgets}
\label{section:provenancewidgets}
\subsection{Design Goals}
We derived seven design goals based on the goals of \edit{prior} provenance visualization tools~\cite{cutler2020trrack,willett2007scented,baudisch2006phosphor} and our own assessment of the \edit{capabilities} we aim to support in \app. Our overarching design goal was to \emph{consistently} achieve the underlying goals for all widgets.

\begin{enumerate}[nosep]
    \item [G1] \textbf{Log User Interactions on UI controls as provenance.} The library should automatically track relevant user interactions with the UI controls (e.g., dragging a slider handle) as provenance.
    \item [G2] \textbf{Compute Aggregated Metrics about Recency and Frequency of Provenance.} The library should process the logged user interactions and compute aggregate summary metrics pertaining to interaction recency and frequency.
    \item [G3] \textbf{Dynamically Overlay Provenance on UI controls.} The library should enhance UI controls with a visual overlay of the aggregate summary metrics and an on-demand temporal evolution of the users' analytic provenance.
    \item [G4] \textbf{Support Action Recovery.} The library should allow navigating historical analysis states and also updating the current state, both programmatically and by interacting with the visual overlays.
    \item [G5] \textbf{Allow Developer Agency.} Application developers should have the flexibility to tune the default tracking and visualization behavior, including being able to disable it completely. The library should provide an API for the same.
    \item [G6] \textbf{Be Framework-Agnostic.} Because multiple frameworks exist, such as Angular, React, and Vue, we derived the goal to design the library to be integrated into any codebase.
    \item [G7] \textbf{Support Meta-Analysis.} The library should support logging and exporting provenance information in a format that is suitable for different kinds of analysis. For example, \app' internal data structure maintains fine-grained logs as well as higher-level computed aggregates.
\end{enumerate}
    
\subsection{Design Process}
As part of our design process, we first reviewed UI controls and then conducted multiple design exercises to decide efficient and consistent visual overlays and associated interactions across all of them.

\subsubsection{UI Controls Review}
We review the structure, layout, and \edit{initial values} of radio buttons~\radiobuttonicon{}, checkboxes~\checkboxicon{}, single sliders~\slidericon{}, range sliders~\rangeslidericon{}, dropdowns~\singleselecticon{}, multiselects~\multiselecticon{}, and input text fields~\inputtexticon{}.

\paragraphHeadingSpace\bpstart{Structure.} All aspects of radio buttons, checkboxes, single sliders, and range sliders are completely visible at all times; whereas, dropdowns, multiselects, and input text fields require an additional click and potential scrolling to bring certain aspects (e.g., options) into focus.

\paragraphHeadingSpace\bpstart{Layouts.} 
Dropdowns, multiselects, and input text fields are oriented \emph{horizontally} with their menus opening vertically (above or below depending on screen position); whereas, radio buttons, checkboxes, single sliders, and range sliders can be oriented \emph{vertically or horizontally}.

\paragraphHeadingSpace\bpstart{Initial/Default \edit{Values}.} 
Radio buttons, checkboxes, dropdowns, multiselects, and input text fields, can have an uninitialized state with \emph{no (null, empty)} selection(s) or value(s); whereas, single sliders and range sliders must always have \emph{at least one} selection. 

\paragraphHeadingSpace\bpstart{Subsequent \edit{Values}.} 
Radio buttons, dropdowns, input text fields, single sliders, and range sliders can have at most \emph{one} selected value; whereas multiselects and checkboxes can have \emph{multiple} selected values.

\paragraphHeadingSpace\bpstart{Interaction Events.} 
Radio buttons, checkboxes, dropdowns, and multiselects are selection-type controls that require the user to \emph{click} to (de)select target options.
Sliders and range sliders require the user to \emph{drag} the handle(s) to \add{or directly \emph{click} on the rail at the} target value(s).
Input text fields \add{generally} require the user to first \emph{type} and then \emph{press the `Enter' key on keyboards} \add{to mark the typing as complete}.
\edit{In \app, any interaction event that modifies the \emph{value (or state)} of a widget is logged and included in its provenance computations; an event that does not modify a widget's value, such as \emph{mouseover} or \emph{keyup} is not logged.}
\add{In addition, clicking on a historical analytic state in the visualization overlays of \app is considered a new interaction and is also appended to the widgets' provenance.}

\subsubsection{Design Exercises and Considerations}
We conducted design exercises to explore \textbf{what} provenance-related information to show, \edit{\textbf{where}, \textbf{how}}, and \textbf{when}.

\paragraphHeadingSpace\bpstart{What \edit{metrics to log as provenance}.} 
\add{We selected two provenance metrics/statistics: frequency and recency of user interactions with widgets (\textbf{G2}). We chose these metrics for their relevance to provenance tracking, intuitive comprehension, effective visual encoding, and broad applicability across various domains.
Furthermore, these metrics can help derive composite metrics such as durations of different widget states and study interaction patterns within and across widgets.}

\paragraphHeadingSpace\bpstart{Where to present provenance.}
We explored on-demand versus always visible visualizations and considered whether they should be juxtaposed against each other, or overlaid or superimposed on the widgets. 
Then, we discussed the trade-offs of having separate overlays against pushing surrounding elements away to accommodate the visual provenance scents. 
\edit{Inspired by Shneiderman's Mantra~\cite{shneiderman2003eyes}, we eventually chose to overlay aggregate views in-place (overview) and temporal views separately on demand considering the level of detail in raw interaction data.}
We designed a tri-state button that would let us toggle between \add{the different} views - default, aggregate, and temporal. 

\begin{figure}[h!]
    \centering
    \includegraphics[width=\linewidth]{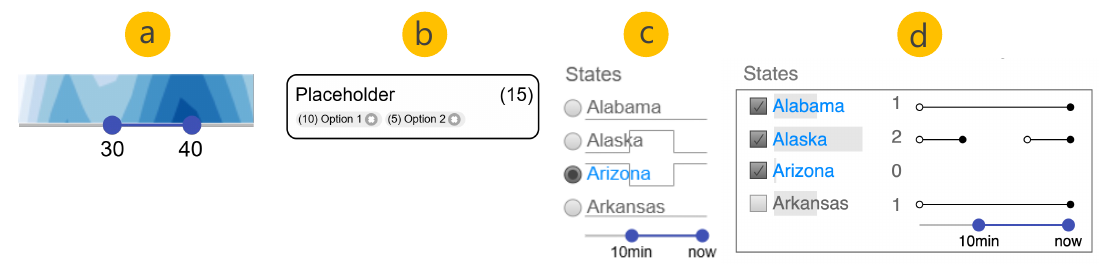}
    \caption{Alternate designs: \add{range slider, input text,} \edit{radio button, checkbox.}}
    \label{fig:alternate-designs}
\end{figure}

\begin{figure*}
    \centering
    \includegraphics[width=\linewidth]{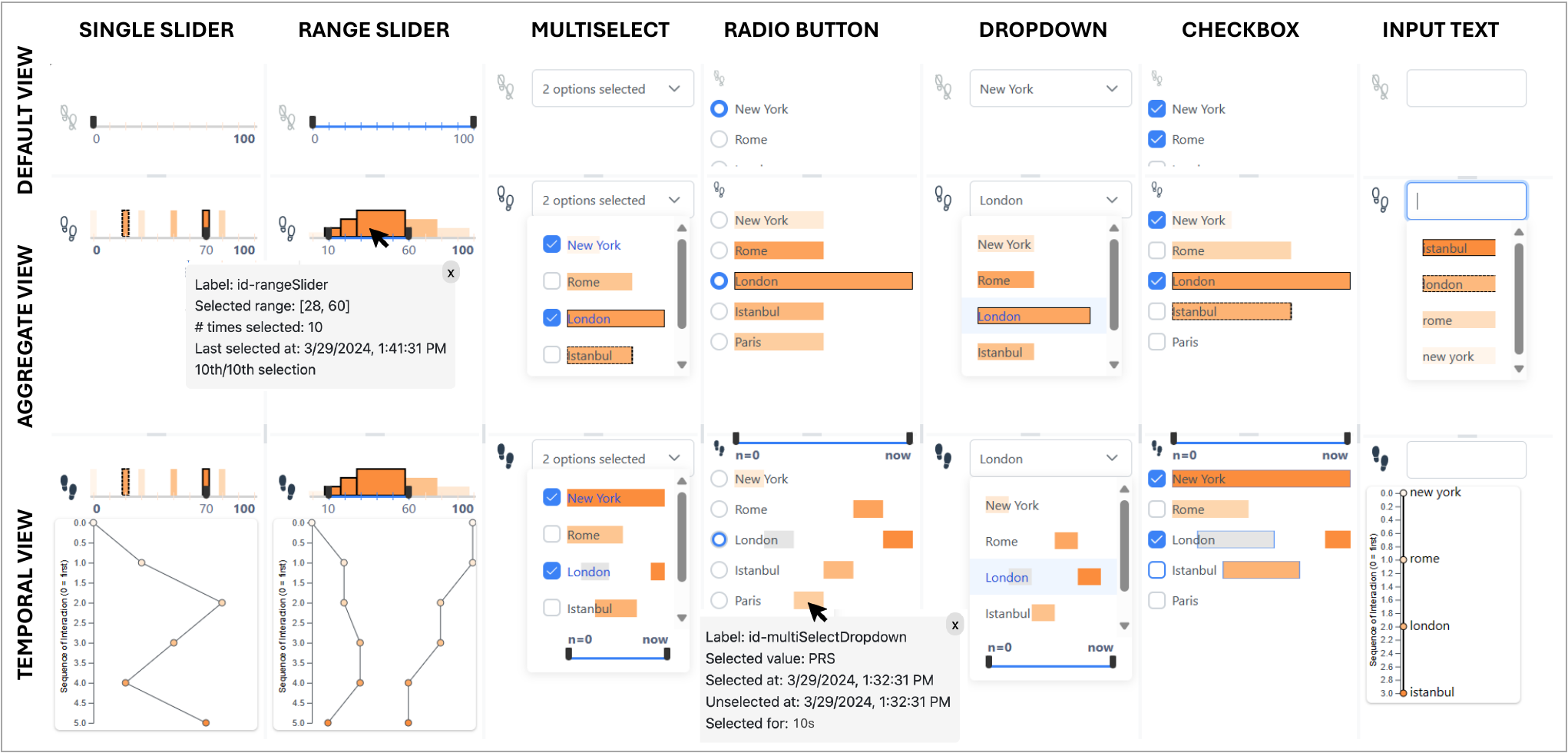}
    \caption{\app: UI controls (single slider, range slider, \edit{multiselect}, radio button, \edit{dropdown}, checkbox, and input text) enhanced with an aggregate summary (Aggregate View) as well as a detailed temporal history (Temporal View) of analytic provenance.}
    \label{fig:provenance-widgets}
    \smallskip
\end{figure*}

\paragraphHeadingSpace\bpstart{How to present provenance.} 
We conducted design exercises to explore candidate visualization and interaction techniques to overlay and interact with the logged provenance information. 
We sketched ideas on draw.io~\cite{drawio} and iterated among co-authors over multiple brainstorming sessions. 
These low-fidelity sketches included considerations for chart types (e.g., bar charts, line charts, and horizon charts), visual encodings (e.g., color, opacity, size), and UI layouts (e.g., panels, overlays).
\add{Keeping in mind our overarching goal of ensuring \emph{consistency}, we selected the \emph{bar} mark and \emph{size, color} encodings to encode frequency and recency information (Figure~\ref{fig:provenance-widgets}).}
Figure~\ref{fig:alternate-designs} shows some of our design considerations for \add{sliders, input texts, radio buttons, and checkboxes.}
\add{For example, we sketched horizon charts in range sliders (Figure~\ref{fig:alternate-designs}\textcolor{orange}{(a)}) and chips in input texts \textcolor{orange}{(b)}, but did not implement them because they did not generalize across all widgets. Similarly, stepped line charts in the temporal view \textcolor{orange}{(c)} seemed occluding and harder to interact with.
We provide all alternate design considerations in supplemental material.}

\begin{figure}[h!]
    \centering
    \includegraphics[width=0.7\linewidth]{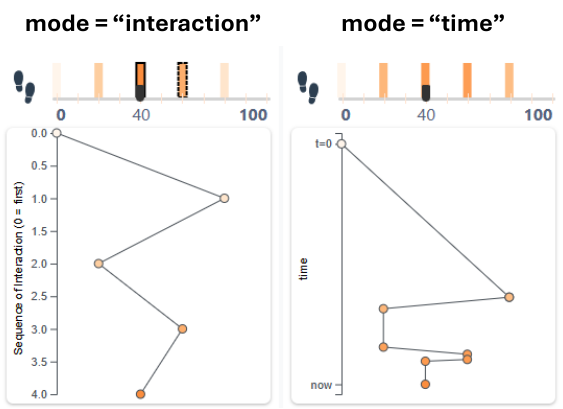}
    \caption{\app: \lstinline[style=TS]{[mode]="interaction"} and \lstinline[style=TS]{[mode]="time"} log interactions every interaction and 1 second (by default), respectively.}
    \label{fig:modes}
\end{figure}

\paragraphHeadingSpace\bpstart{When to log provenance}. We considered two kinds of logging frequencies -- \emph{interaction-based}, which captures every user interaction when it occurs, and \emph{time-based}, which captures snapshots at specific intervals.
Finding utility in both, we chose to support both modes (\textbf{G1}\add{, Figure~\ref{fig:modes}}).

\subsection{Widget Design}

Figure~\ref{fig:provenance-widgets} shows our chosen designs for the provenance overlays. The default view of each widget is enhanced by the \textbf{Aggregate View} (aggregate summary) and a \textbf{Temporal View} (detailed history) (\textbf{G3}).

When the widget has not been interacted with (i.e., there is no logged provenance), a disabled footprint icon-button \disabledmodeicon{} is placed next to the UI control. 
When the widget is interacted with for the first time, this icon-button is enabled, and the widget switches to the Aggregate view \aggmodeicon{}, which overlays aggregate provenance information.
Clicking the footprint icon toggles between this Aggregate view \aggmodeicon{} and the Temporal view \tempmodeicon{}, that overlays the temporal history of provenance.

\subsubsection{Single Slider~\slidericon{}, Range Slider~\rangeslidericon{}}

\paragraphHeadingSpace\bpstart{Aggregate View.} We designed a bar chart that shows previously selected values (slider) or ranges of values (range slider). 
The frequency of a selection is encoded by height, and the recency of a selection is encoded by color. 
Taller, darker bars indicate more frequent and recent interactions, respectively.
This bar chart is positioned directly above the slider, as in Scented Widgets~\cite{willett2007scented}. 
Hovering a bar shows a tooltip with the value, timestamp, frequency, and recency of the selection. Clicking a bar updates the slider to the selected value or range.

\paragraphHeadingSpace\bpstart{Temporal View.} 
To visualize the temporal evolution, we designed a popover that is overlaid above or below the slider. 
Within this popover, we designed a line chart where time is measured along the y-axis, and the slider itself serves as the x-axis.
This line chart has one line for a single slider and two lines for a range slider (one for each handle). 
The line(s) have circular points that represent the exact selections made over time. 
These points are also colored by the recency of the selections. 
Hovering a point shows a tooltip with the corresponding value and time of the selection. 
In addition, clicking a point updates the slider to the selected value or range (\textbf{G4}). 
Lastly, to facilitate navigation, the y-axis can also be brushed to zoom in on more granular, specific time ranges.

\subsubsection{Dropdown~\singleselecticon{}, Multiselect~\multiselecticon{}, \\Radio Button~\radiobuttonicon{}, Checkbox~\checkboxicon{}}

We collectively refer to dropdowns, multiselects, radio buttons, and checkboxes as \emph{selection-type} widgets as they all have a similar visual and interaction design for interacting with provenance information.

\paragraphHeadingSpace\bpstart{Aggregate View.} We designed a bar chart and placed it under the options list. 
An option's selection frequency is encoded by the length of the bar underneath it and the recency is encoded by color.
Longer and darker bars indicate higher frequency, recency, respectively.
Hovering a bar shows a tooltip with the value, timestamp, frequency, and recency of the selection.
Clicking a bar updates the option's selection. 

\paragraphHeadingSpace\bpstart{Temporal View.} 
To visualize the temporal evolution, we directly modify the aggregate bar chart unlike that in sliders, where we create a new popover.
Each bar represents the time range during which the option was selected. 
The length of the bar represents the duration of the selection, and the color represents the recency. 
Longer, darker bars indicate higher frequency, recency, respectively.
Hovering a bar shows a tooltip with the corresponding time range.
Clicking a bar selects the corresponding option, along with other options that were selected at that point in time. 
Lastly, to facilitate navigation, there is a horizontal range slider to zoom in on more granular, specific time ranges. 

\subsubsection{Input Text~\inputtexticon{}}

\paragraphHeadingSpace\bpstart{Aggregate View.} 
We utilize a dropdown list of previously entered values and visualize provenance as a bar chart underneath each list item.
The frequency of an input value is encoded by the length of the bar underneath it, and the recency of a selection is encoded by color. 
Longer, darker bars indicate higher frequency, recency, respectively.
Hovering a list item shows a tooltip with the corresponding timestamp, frequency, and recency of the input value.
Clicking a list item updates the text input selection to the corresponding value.

\paragraphHeadingSpace\bpstart{Temporal View.} 
To visualize the temporal evolution, we designed a popover that is overlaid above or below the text input. 
Within this popover, we designed a vertical timeline chart that shows what text input was searched and when. 
This timeline has circular points that represent the exact search inputs made over time. 
These points are also colored by the recency of the input searches. 
Hovering a point shows a tooltip with the corresponding timestamp of input.
Clicking a point updates the current text input selection to the corresponding value.

\subsection{Architecture}

We broadly define the architecture of \app using \textbf{MVC} (\model{}-\view{}-\controller{}), a software design pattern commonly used to develop GUIs (Figure~\ref{fig:teaser}). We describe these as follows:

\paragraphHeadingSpace\bpstart{\view{}}: \textit{What the user interacts with} - The \view{} handles all concerns related to the appearance of the widgets, including the base widgets and the overlaid provenance. Internally, we define it almost entirely with Angular templates (HTML) and CSS.

\paragraphHeadingSpace\bpstart{\controller{}}: \textit{What the developer interacts with} - The \controller{} serves as a hub between the developer, the \view{}, and the \model{}. 
Essentially, it wraps the base widget and exposes all of its properties and events, in addition to the \app API. It passes on all the base widgets' properties to the \view{} templates, and intercepts all incoming events before re-emitting them for the developers. If not frozen, it relays these events and all provenance-related properties to the \model{}.

\paragraphHeadingSpace\bpstart{\model{}}: \textit{What we interact with} - The \model{} stores the raw interaction data received from the \controller{}, and uses it to compute frequency (how many times a value was input) and recency (how recently a value was input). Once the provenance is updated, the \model{} can emit it via the \controller{} as an event for the developers to subscribe to. Then, if visualization is enabled, it updates the \view{} with aggregated summaries of frequency and recency (Aggregate View) or raw temporal history (Temporal View) depending on the active mode.

\subsection{Implementation} \label{Implementation}

\app is implemented using Angular~\cite{angular} with an extensible API to support flexibility across different systems. To ensure portability across frameworks (e.g., Vue~\cite{vue}, React~\cite{react}), we leverage the WebComponents API (\textbf{G6}).
Below we describe the main properties (attributes) and events available in the \app API.

\begin{enumerate}[nosep, leftmargin=0.4cm]
    \item \textbf{provenance}: The information recorded and computed by the widget to visualize the provenance of interactions. While each widget has a unique provenance structure, all of them record interactions as an array of objects with the selected/input value and the timestamp of the interaction. This property can be used to initialize, restore, modify, and export (\textbf{G7}) the provenance of the widget.
    \item \textbf{provenanceChange}: An event that is triggered whenever the user interacts with the widget \edit{such that its \emph{value} (and hence provenance) changes. For example, \emph{clicking} a radiobutton option or \emph{dragging} a range slider handle constitute a valid event; however, \emph{keyup} or \emph{mouseover} events do not contribute to the provenance.}
    \item \textbf{mode}: This property configures the provenance logging frequency (Figure~\ref{fig:modes}). When `mode' is set to \emph{``interaction''}, the widget logs every user interaction and accordingly recomputes provenance metrics and updates the subsequent visualizations.
    When `mode' is set to \emph{``time''}, the widget logs interactions every `t' seconds (t=1 second by default) and accordingly updates everything downstream.
    \item \textbf{freeze}: A property to stop logging interactions with the widget. When `freeze' is set to true, the widget will not record any new interactions, and existing visualizations will not be updated. When `freeze' is set to false, the widget will continue recording and visualizing the provenance from the last recorded interaction (\textbf{G4}).
    \item \textbf{visualize}: A boolean property to toggle the visibility of the provenance overlays. This property can be used (along with `freeze'=true) to completely disable provenance in the widget (\textbf{G4, G5}).
    \item \textbf{data-label}: An attribute to pass additional context to the widget. This is displayed as ``label'' in the widgets' tooltips.
\end{enumerate}

\begin{lstlisting}[style=TS]
provenance?: SliderProvenance | InputTextProvenance | Provenance
mode?: "interaction" | "time"
\end{lstlisting}
\begin{lstlisting}[
style=Angular, 
% caption={Common Provenance Widgets Attributes}, 
label={lst:provenance-widget-generic}]
<provenance-{slider,dropdown,multiselect,radiobutton,checkbox,inputtext}
    [(provenance)]="provenance" [mode]="mode"
    [visualize]="true" [freeze]="false"
    [attr.data-label]="`label'"/>
\end{lstlisting}

\paragraphHeadingSpace\bpstart{Understanding code snippets and notations.} We describe the \app API using TypeScript and Angular's data binding syntax, which can be categorized based on data flow:

\begin{enumerate}[nosep]
    \item \emph{From source to view (property binding).} \lstinline[style=TS]{[property]="expression"} binds the value from the expression to the property. Can be also used to bind class and style properties, and \verb|data-*| attributes.
    \item \emph{From view to source (event binding).}
    \lstinline[style=TS, mathescape]{(event)="function($\mbox{\textdollar}$event)"} executes the bound \verb|function| with the \verb|$event| object emitted by the \verb|event|.
    \item \emph{In both ways (two-way binding).} \lstinline[style=TS]{[(property)]="expression"} binds the value from the expression to the property, and vice versa. It is syntactic sugar for combining property and event binding. For example, \verb|[(provenance)]| is syntactic sugar for \verb|[provenance]| and \verb|(provenanceChange)|.
\end{enumerate}

\subsubsection{Single Slider~~\slidericon{}, Range Slider~~\rangeslidericon{}}
A slider allows users to select a numeric value from a given range. Traditionally defined as {\fontfamily{cmr}\selectfont<input type="range">} in HTML, these elements only allow for a single value to be selected. In addition to the traditional sliders, \app also provides Range Sliders, which permit selection of a range of values.

\begin{lstlisting}[style=TS]
value: number = 0
highValue?: number = 0 // Omit for Single Slider
handleChange(event: ChangeContext) {
    value = event.value
    highValue = event.highValue }
options: Options = { floor: 0, ceil: 100, step: 1 }
\end{lstlisting}
\begin{lstlisting}[style=Angular, 
% caption={Range Slider (and Single Slider)}, 
label={lst:rangeslider}]
<provenance-slider [options]="options"
    [value]="value" [highValue]="highValue"
    (selectedChange)="handleChange($event)" />
\end{lstlisting}

The Slider widget extends \href{https://angular-slider.github.io/ngx-slider/api-docs/classes/SliderComponent.html}{SliderComponent from @angular-slider/ngx-slider}, and exposes an additional {\fontfamily{cmr}\selectfont selectedChange} event which is triggered at the end of user's interaction with the slider.

\subsubsection{Text Input~~\inputtexticon{}}

A Text Input allows users to enter values comprising of text, numbers, and symbols. Traditionally defined as {\fontfamily{cmr}\selectfont<input type="text">} in HTML, these elements create a basic single-line text input field.

\begin{lstlisting}[style=TS]
value: string = ''
\end{lstlisting}
\begin{lstlisting}[style=Angular,  label={lst:inputtext}]
<provenance-inputtext [(value)]="value" />
\end{lstlisting}

The Text Input widget extends \href{https://www.primefaces.org/primeng-v15-lts/autocomplete}{PrimeNG's AutoComplete} component and exposes an additional {\fontfamily{cmr}\selectfont valueChange} event which is triggered when the input value changes.

\subsubsection{Dropdown~~\singleselecticon{}}

A Dropdown allows users to select a single value from a list of options. In HTML, these elements are defined using the {\fontfamily{cmr}\selectfont<select>} tag and a list of {\fontfamily{cmr}\selectfont<option>} tags nested within it.

\begin{lstlisting}[style=TS]
type Option = { label: string, value: string }
options: Option[] = []
selected?: Option
\end{lstlisting}
\begin{lstlisting}[style=Angular, 
% caption={Dropdown}, 
label={lst:dropdown}]
<provenance-dropdown [options]="options"
    optionLabel="label" dataKey="value"
    [(selected)]="selected" />
\end{lstlisting}

In \app, the Dropdown widget extends \href{https://www.primefaces.org/primeng-v15-lts/dropdown}{PrimeNG's Dropdown} component and exposes its {\fontfamily{cmr}\selectfont options} attribute to allow developers to provide their list of options.

\subsubsection{Multiselect~~\multiselecticon{}}

A Multiselect input allows users to select multiple values from a list of options. In HTML, these are defined in the same way as Dropdowns, but with the {\fontfamily{cmr}\selectfont multiple} attribute set to true on the {\fontfamily{cmr}\selectfont<select>} tag. However, unlike Dropdowns, a Multiselect input renders a list of scrollable options and requires the user to hold down the control key while clicking to select multiple options.

\begin{lstlisting}[style=TS]
selected?: Option[]
\end{lstlisting}
\begin{lstlisting}[style=Angular, 
% caption={Multiselect}, 
label={lst:multiselect}]
<provenance-multiselect [options]="options"
    optionLabel="label" dataKey="value"
    [(selected)]="selected" />
\end{lstlisting}

In \app, the Multiselect widget extends \href{https://www.primefaces.org/primeng-v15-lts/multiselect}{PrimeNG's MultiSelect} component and exposes its {\fontfamily{cmr}\selectfont options} attribute to allow developers to provide their list of options. Unlike a traditional multiselect input, the \app \ Multiselect widget renders in a Dropdown-like manner and does not require users to hold down any keys to select multiple options.

\subsubsection{Radio Button~~\radiobuttonicon{}}

A Radiobutton allows users to select a single value from a list of options. In HTML, these are defined using the {\fontfamily{cmr}\selectfont<input type="radio">} tag, and all radio buttons with the same {\fontfamily{cmr}\selectfont name} attribute are grouped together.
\begin{lstlisting}[style=TS]
selected?: string
\end{lstlisting}
\begin{lstlisting}[style=Angular, 
% caption={Radio button}, 
label={lst:radiobutton}]
<provenance-radiobutton [data]="options" 
    [(selected)]="selected" />
\end{lstlisting}

In \app, the Radio Button widget extends \href{https://www.primefaces.org/primeng-v15-lts/radiobutton}{PrimeNG's RadioButton} component. However, unlike traditional Radio Buttons, the \app \ Radio Button widget represents a group of vertically aligned self-contained radio buttons. It exposes a {\fontfamily{cmr}\selectfont data} attribute, which allows developers to provide their list of options instead of having to define each radio button individually.

\subsubsection{Checkbox~~\checkboxicon{}}

A Checkbox allows users to select or deselect a single value. Checkboxes can be standalone, or grouped together with the same {\fontfamily{cmr}\selectfont name} attribute. In HTML, these are defined using the {\fontfamily{cmr}\selectfont<input type="checkbox">} tag.

\begin{lstlisting}[style=TS]
selected ?: string[]
\end{lstlisting}
\begin{lstlisting}[style=Angular, 
% caption={Checkbox}, 
label={lst:checkbox}]
<provenance-checkbox [data]="options" 
    [(selected)]="selected" />
\end{lstlisting}

In \app, the Checkbox widget extends \href{https://www.primefaces.org/primeng-v15-lts/checkbox}{PrimeNG's Checkbox} component. Like the Radio Button widget, the \app \ Checkbox widget exposes a {\fontfamily{cmr}\selectfont data} attribute, which allows developers to provide their list of options.
All \textit{selection-type} widgets expose a {\fontfamily{cmr}\selectfont selected} attribute, that allows developers to provide an initial selection or override the current selection, and a {\fontfamily{cmr}\selectfont selectedChange} event, triggered when the selection changes.

\section{Evaluation}
\label{section:evaluation}

\begin{figure}
    \centering
    \includegraphics[width=0.85\linewidth]{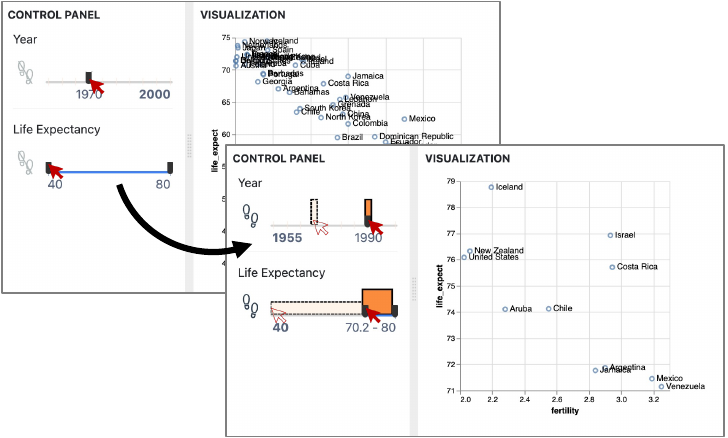}
    \caption{Visualize Visualization Specifications and Transformation Interactions on \app.}
    \label{fig:widget-to-vis}
\end{figure}

\subsection{Example Usage Scenarios}

\bpstart{Track Data Transformation and Visualization Specification Interactions (Widget to Visualization).}
\app can help users track what charts they make (visualization specification) and what filters they apply (data transformations).
Consider Figure~\ref{fig:widget-to-vis} that shows a scatterplot visualization of two attributes: ``Year'' and ``Life Expectancy'' along with corresponding single slider and range slider \app.
As the user drags the slider handle(s): ``Year'': 1970 $\rightarrow$ 1990 and ``Life Expectancy'': [40, 80] $\rightarrow$ [70.2, 80], the scatterplot updates and also the orange provenance overlays become visible.
In this way, the user can utilize \app to track already explored data ranges, potentially informing subsequent explorations.

\begin{lstlisting}[
    style=TS,
    label={lst:widget-to-vis-html}
    % frame=bottomline,
    % caption={Interaction with slider updated embedded Vega visualization}
]
const { view } = await embed("spec.vg.json", ...)
const slider = document.createElement("web-provenance-slider");
slider.value = 0;
slider.addEventListener("selectedChange", e => {
    view.signal("slider", e.detail.value).runAsync() })
\end{lstlisting}

In the above listing, the developer consumes \app as Web Components and binds properties and events in JavaScript. They subscribe to \lstinline[style=TS]{"selectedChange"} to update the embedded Vega chart~\cite{satyanarayan2016vega}.

\paragraphHeadingSpace\bpstart{Track Direct Manipulation-based Interactions in Visualizations (Visualization to Widget).}
\label{subsubsection:vistowidget}
Because not all user interactions happen via UI controls, \app can be externally updated when user interactions happen elsewhere, e.g., in the visualization.

\begin{figure}
\smallskip
    \centering
    \includegraphics[width=0.85\linewidth]{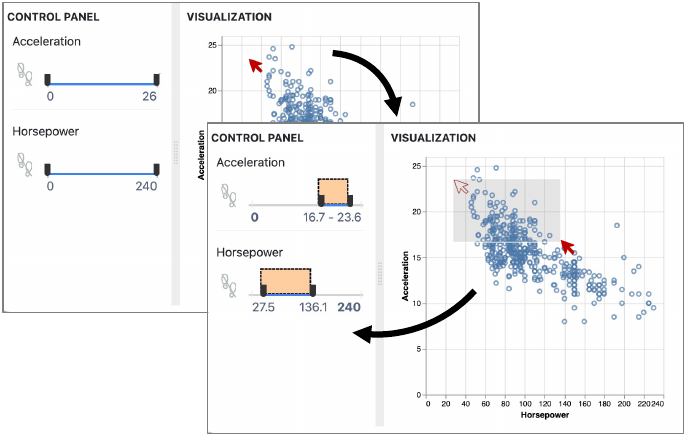}
    
    \caption{Track Direct Manipulation-based Interactions in Visualizations.}
    \label{fig:vis-to-widget}
\end{figure}

Consider Figure~\ref{fig:vis-to-widget} that shows a scatterplot visualization of two attributes and corresponding range slider \app: ``Acceleration'' and ``Horsepower''. As the user performs a brush interaction in the visualization, selecting a subset of points within a specific range (``Horsepower'': [27.5, 136.1] and ``Acceleration'': [16.7, 23.6]), the corresponding range sliders can update to show this range.
In this way, the user can utilize \app to track what data ranges they have already explored, potentially informing subsequent explorations.

\begin{lstlisting}[style=TS]
visBrushed(brush_extent) {
    acc.provenance["revalidate"] = true
    acc.provenance["data"] = [{ ..., "value": brush_extent }] // [16.7. 23.6]
}
\end{lstlisting}
\begin{lstlisting}[
style=Angular, 
% caption={Brushing in Visualization updates Range Slider}, 
label={lst:vis-to-widget-html}]
<provenance-slider [(provenance)]="acc.provenance" />
\end{lstlisting}

In the above listing, the developer subscribes to the visualization's brush event via \lstinline[style=TS]{"visBrushed()"} and updates the \lstinline[style=TS]{"provenance"} of the ``Acceleration'' (\lstinline[style=TS]{"acc"}) range slider.

\subsection{Replicating Prior Work}
We utilize \app to replicate three prior works in improving social navigation cues (Scented Widgets~\cite{willett2007scented}), explaining transitions in the user interface (Phosphor Objects~\cite{baudisch2006phosphor}), and dynamic querying-based~\cite{shneiderman1994dynamic,williamson1992dynamic} or direct manipulation-based~\cite{heer2008generalized} interactions with a visualization system (Dynamic Query Widgets).

\subsubsection{Scented Widgets.} 
Scented Widgets~\cite{willett2007scented} are graphical user interface controls enhanced with embedded visualizations that facilitate navigation in information spaces. 
For example, each option in a radio button includes a visual scent of the number of times it has been used across multiple users.

\app can be configured to recreate these widgets by showing static information about (2) social navigation, e.g., number of times each radio button option was chosen across multiple users and when (Figure~\ref{fig:scented-widgets-phosphor-objects-dynamic-query-widgets}A-shades of orange) or (2) data distribution, e.g., distribution of values for that column in the underlying dataset (Figure~\ref{fig:scented-widgets-phosphor-objects-dynamic-query-widgets}A-blue).
To realize the range slider in Figure~\ref{fig:scented-widgets-phosphor-objects-dynamic-query-widgets}A-shades of orange, the developer can program the widget in the following way:

\begin{lstlisting}[style=TS]
historical_usage_logs = {
  "revalidate": true,
  "data": [{"value": [100, 160], "timestamp": _},
           {"value": [100, 160], "timestamp": _},
           {"value": [160, 200], "timestamp": _}, ...]
}
\end{lstlisting}
\begin{lstlisting}[
style=Angular, 
% caption={Recreating a Range Slider Scented Widget},
label={lst:scented-widget-rangeslider-html}]
<provenance-slider [freeze]="true"
    [(provenance)]="historical_usage_logs" />
\end{lstlisting}

In the above listing, the developer passes the \lstinline[style=TS]{"historical_usage_logs"} information in the format of interaction logs, which is then mapped to the \lstinline[style=TS]{[(provenance)]} property. The \lstinline[style=TS]{[freeze]="true"} property will ensure the widgets don't update in real-time with user interactions.

\subsubsection{Phosphor Objects.} 
Phosphor objects~\cite{baudisch2006phosphor} employ a \emph{phosphor transition} as a transition that (1) shows the outcome of the change instantly via an afterglow effect and (2) also explains the change in retrospect using a diagrammatic depiction. 
For example, manipulating a phosphor slider leaves an afterglow that illustrates how and from where the knob moved.
Users who already understand the transition can continue interacting without delay, while those who are inexperienced or may have been distracted can take time to view the effects at their own pace. 

\begin{figure*}[t]
    \centering
    \includegraphics[width=0.95\linewidth]{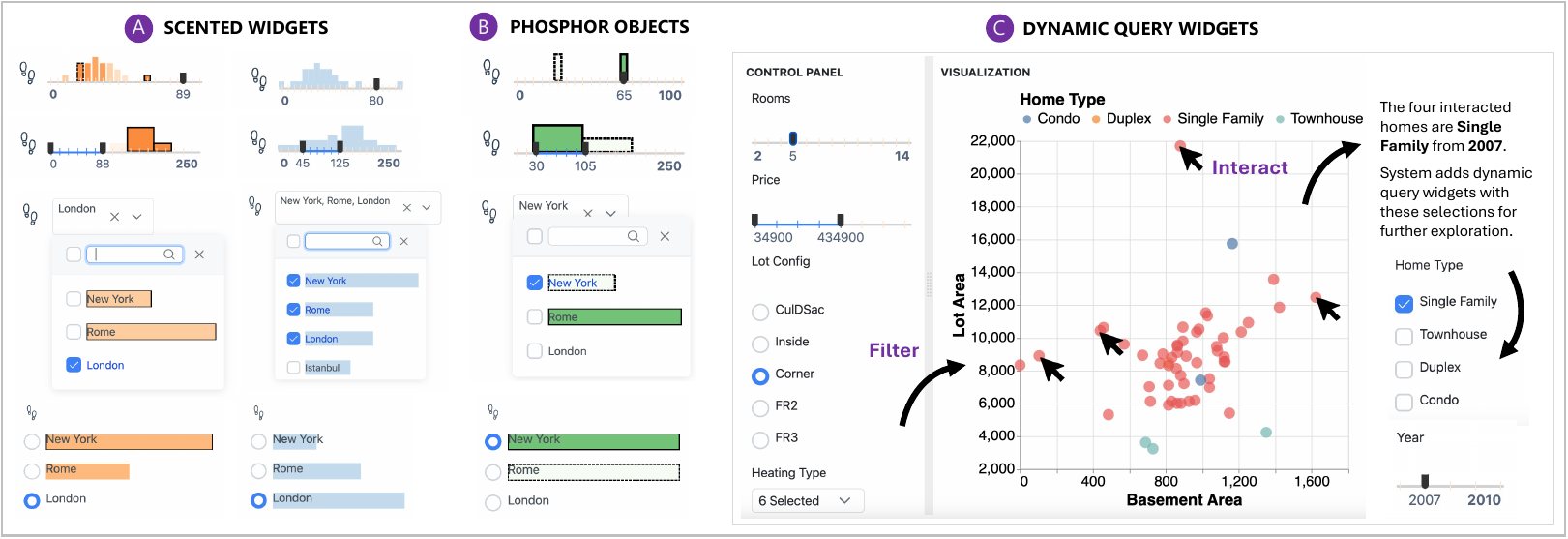}
    \caption{\app can be configured to (re)create the core functionalities of (a) Scented Widgets, (b) Phosphor Objects, and (c) Dynamic Query Widgets. 
    Scented Widgets enhance UI controls via embedded visualizations of some pre-computed metric, e.g., visit frequency and recency (in shades of orange) or data distribution (in blue) to facilitate navigation. 
    Phosphor objects track user interactions with UI controls in real-time and leave visual scents of the most recent (dark green) and second most recent (light green) interaction. 
    Dynamic Query Widgets are UI controls that continuously update a visualization and/or its underlying data as the user adjusts them. For example, \app can facilitate creating a dynamic query~\cite{shneiderman1994dynamic} to lookup affordable (``Price'' < \$500k) houses with five ``Rooms'' and ``Lot Config''=\emph{Corner} and then update a visualization to see the query result. Alternatively, these widgets can also be created on the fly, e.g., if a user interacts with ``Home Type''=\emph{Single Family} and ``Year''=\emph{2007} houses, the system can add new query widgets for ``Home Type'' and ``Year'' to generalize the user's selection~\cite{heer2008generalized} and facilitate future exploration.}
    \label{fig:scented-widgets-phosphor-objects-dynamic-query-widgets}
\end{figure*}

\app can be configured to recreate Phosphor objects by limiting the \emph{recency of interaction} mapping to the color encoding channel to just include the two most recent interactions (the current and the previous interaction). 
That way, every interaction will leave behind a single visual trace (e.g., light green bar) corresponding to the previous \edit{value}.
To realize the single slider in Figure~\ref{fig:scented-widgets-phosphor-objects-dynamic-query-widgets}B, the developer can program the widget in the following way:

\begin{lstlisting}[style=TS]
widgetUpdated() {
  if (provenance) provenance = { 
    "revalidate": true,
    "data": provenance["data"].slice(-2) }
}
\end{lstlisting}
\begin{lstlisting}[
style=Angular, 
% caption={Recreating a Single Slider Phosphor Object}, 
label={lst:phosphor-object-singleslider-html}]
<provenance-slider [(provenance)]="provenance"
    (provenanceChange)="widgetUpdated()" />
\end{lstlisting}

In the above listing, when a widget is interacted with (\lstinline[style=TS]{"widgetUpdated()"}), the developer modifies the \lstinline[style=TS]{"provenance"} array by slicing it to only keep the two most recent interactions and then commanding the library to \lstinline[style=TS]{revalidate} and recompute its internal model and update the view.

\subsubsection{Dynamic Query Widgets.} 

\bpstart{Dynamic querying.} Shneidermann~\cite{shneiderman1994dynamic} introduced the notion of dynamic queries to continuously update the data that is filtered from the database and visualized. These queries ideally work instantly as the user adjusts UI controls such as sliders or dropdowns to form simple queries or to find patterns or exceptions. 
Williamson~et~al.~\cite{williamson1992dynamic} then evaluated this approach in a real-estate system called HomeFinder.

Figure~\ref{fig:scented-widgets-phosphor-objects-dynamic-query-widgets}C shows how \app can create HomeFinder. The developer can program the ``Rooms'' single slider as follows:

\begin{lstlisting}[style=TS]
// Apply the filter and update the visualization
widgetUpdated(model) {}
\end{lstlisting}
\begin{lstlisting}[
style=Angular, 
% caption={Creating a Dynamic Query Widget (sans provenance tracking)}, 
label={lst:dynamic-query-widget-singleslider-html}]
<provenance-slider [visualize]="false" [freeze]="true"
    (selectedChange)="widgetUpdated($event.value)"/>
\end{lstlisting}

In the above listing, the widget is initialized with \lstinline[style=TS]{[freeze]}=``true'' and \lstinline[style=TS]{[visualize]}=``false'', which disables logging and hides any overlays. When a widget is interacted with (\lstinline[style=TS]{"widgetUpdated()"}), the developer can access the new model, filter the data, and update the visualization.

\paragraphHeadingSpace\bpstart{Direct manipulation.} Heer et al.~\cite{heer2008generalized} introduced direct manipulation techniques that couple declarative selection queries with a query relaxation engine, enabling users to interactively generalize their selections using dynamically generated query widgets. 
For example, if a user's selections on a housing dataset only include ``Home Type''=\emph{Single Family} and ``Year''=\emph{2007}, then two dynamic query widgets are created: a checkbox group for ``Home Type'' with the \emph{Single Family} option checked; and a single slider for ``Year'', preset to \emph{2007}.

Figure~\ref{fig:scented-widgets-phosphor-objects-dynamic-query-widgets}C shows how \app can support dynamic query widgets created via direct manipulation. To realize the ``Year'' single slider, the developer can program the widget as follows:

\begin{lstlisting}[style=TS]
selectedYear = 2007;
showWidget = true;
\end{lstlisting}
\begin{lstlisting}[
style=Angular, 
% caption={Creating a Dynamic Query Widget (sans provenance tracking)}, 
label={lst:direct-manipulation-widget-singleslider-html}]
<provenance-slider *ngIf="showWidget"
    [visualize]="false" [freeze]="true"
    [selected]="selectedYear" />
\end{lstlisting}

In the above listing, the widget is created (or made visible) by \lstinline[style=TS]{*ngIf}=``showWidget'' and initialized with \lstinline[style=TS]{[selected]}=``selectedYear'' (the output of the generalized selection algorithm). The other properties, \lstinline[style=TS]{[freeze]} and \lstinline[style=TS]{[visualize]} are still set to ``true'' and ``false'', respectively.

\subsection{Cognitive Dimensions of Notation}

We self-assess our library from a developer standpoint based on the Cognitive Dimensions of Notation~\cite{blackwell2001cognitive}, a framework of heuristics commonly used to assess the effectiveness of notational systems (e.g., visualization grammars and toolkits). 
Of the 14 cognitive dimensions, we select a relevant subset for comparing our work with existing tools.

\paragraphHeadingSpace\bpstart{Consistency:} \emph{Similar semantics are expressed in similar syntactic forms} -- 
\add{\app exposes a common set of provenance-related properties and events, which behave consistently across all underlying widgets (described in Section~\ref{Implementation}). The notation is also consistent with the base libraries it inherits from, as well as consistent across different JavaScript frameworks when used as Web Components.}

\paragraphHeadingSpace\bpstart{Diffuseness:} \emph{Verbosity of language} and\bpstart{\add{Hard Mental Operations:}} \add{\emph{high demand on cognitive resources}} -- Since provenance is built into the widgets, developers can directly use components from the underlying libraries to create provenance-aware widgets.
Even advanced use cases such as persisting, restoring, or modifying the provenance \edit{only require minimal code} and \add{cognitive demand.} This is in contrast to existing provenance systems such as Trrack and TrrackVis~\cite{cutler2020trrack}, which require developers to set up states, actions, event listeners, and other components to capture and visualize provenance.

\paragraphHeadingSpace\bpstart{Viscosity:} \emph{Difficulty of making changes} -- The widgets have a low viscosity for primitive attributes, but a high viscosity for complex attributes. For example, developers can easily add labels and toggle provenance tracking and visualization. However, changing the options and provenance data structures requires more effort.

\subsection{Expert Case Studies}
In this section, we present case studies with developers who utilized \app to create custom applications from the ground up. Our aim was to evaluate how effective \app is for building provenance-focused visual data analysis systems, as well as to understand developers' experiences working with it, including installation, configuration, and customization.

\paragraphHeadingSpace\bpstart{Participants.} We recruited four developers (\user{1-4}) well versed in front-end web development and visual data analysis.
Demographically, they were men (2) and women (2) in the 18-24 (1) and 25-34 (3) age groups.

\paragraphHeadingSpace\bpstart{Task.} We tasked participants to:
\begin{quoting}[leftmargin=18pt, rightmargin=18pt, vskip=0pt]
    \emph{Develop a ``Pokemon Explorer'' visualization system for a Pokemon Fan Club, to help member fans visually explore Pokemon names and stats to pick their dream team.
    The visualization system should consist of a visualization, and UI controls, that help specify the visualization (e.g., map variables to visual encodings) and/or beautify it (e.g., modify font styles and color schemes).
    The Club wishes to track fans' interaction behaviors as they explore the data, hence you must use \app as your UI controls to help track and visualize each user's analytic provenance.}
    \emph{In addition, try to capture relevant user interactions from other, non-\app places in your application, and manually update \app. For example, brushing within a scatterplot visualization should log the brushed extents on either axis and append them to the provenance data structures of the corresponding attribute filters.}
\end{quoting}

\paragraphHeadingSpace\bpstart{Dataset.} We used a dataset of 802 pokemon~\cite{serebii} comprising
nine quantitative variables (\emph{Height\_m}, \emph{Weight\_kg}, \emph{HP}, \emph{Speed}, \emph{Attack}, \emph{Special Attack}, \emph{Defense}, \emph{Special Defense}, \emph{Happiness}), 
five nominal variables (\emph{Classification}, \emph{Name}, \emph{Primary Type}, \emph{Secondary Type}, \emph{Is Legendary}), 
and two ordinal variables (\emph{Pokedex Number}, \emph{Generation}). This variety of variables enables developers to use different widgets.

\begin{figure*}[t]
    \centering
    \includegraphics[width=0.85\linewidth]{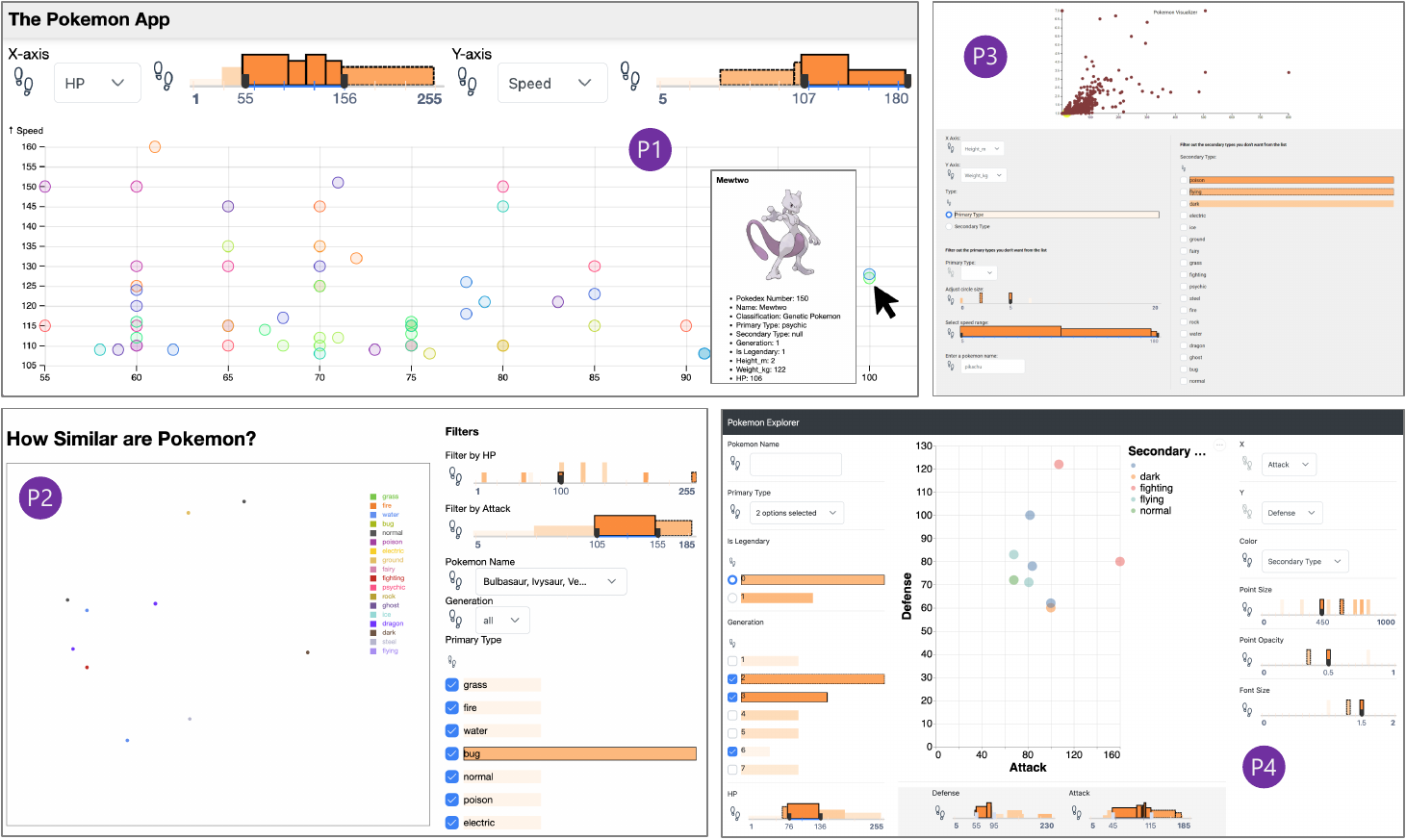}
    \caption{Pokemon Explorer applications as developed by our developer participants. All participants created a scatter plot-based visualization system using \app to apply filters (\user{1,2,3,4}), specify visual encodings: \emph{xy} (\user{1, 2, 4}), \emph{color} (\user{1, 4}), and/or adjust styling (\user{3, 4}). }
    \label{fig:all-applications}
    \smallskip
\end{figure*}

\paragraphHeadingSpace\bpstart{Logistics.}
We first conducted a 30-minute onboarding interview over Zoom, during which we sought consent from participants and introduced them to \app and the study task. We also asked them their preference between the Angular components and the WebComponent versions of the library. Accordingly, we shared a Github repository with them that included installation instructions, API documentation, task instructions, and starter code.

Next, we gave participants up to one week to complete the task. During this week, we asked them to document their experience (e.g. bugs, happy moments) working with the widgets in a {\fontfamily{cmr}\selectfont FEEDBACK.md} file. If and when stuck, we asked participants to create GitHub issues or directly email the study administrators.

Finally, we conducted a 30-minute debriefing interview wherein we reviewed the participants' visualization systems, source code, and feedback notes.
We compensated each participant with a \$25 gift card.

\paragraphHeadingSpace\bpstart{Analysis.}
We manually transcribed the audio recordings and feedback, divided them into smaller sections, and applied open coding~\cite{boyatzis1998transforming}, specifically, constant comparison and theoretical sampling~\cite{strauss1998basics}.
\edit{Next, we briefly describe the participants' applications, development experience, and feedback on the widgets, including future enhancements.}

\subsubsection{Developed Applications}
Figure~\ref{fig:all-applications} shows four applications developed by our participants using Angular (\user{1,4}) and Web components (\user{2,3}).
All participants created a scatter plot-based system using \app to apply filters (\user{1,2,3,4}), specify visual encodings: \emph{xy} (\user{1, 2, 4}), \emph{color} (\user{1, 4}), and/or adjust styling (\user{3, 4}). 
\user{2}'s scatterplot visualized the output of a UMAP~\cite{mcinnes2018umap} dimensionality reduction algorithm that \edit{groups more similar pokemon to be closer to each other}.
\user{4} wanted to \edit{be able to} export the visualizations, hence they provided additional options to configure the font size, point size, and point opacity.
Only \user{4} \edit{attempted the bonus task, } to capture user interactions externally (via brushing in the visualization) and manually update the provenance on the relevant widget.

\subsubsection{Developer Experience}

\add{\bpstart{General Feedback.}} \add{All four developers found \app to be useful, commending its built-in capability to track and visualize provenance. 
\user{1} said, \emph{``I was initially very surprised and actually very excited like, wow, it's very well-made, doesn't really break. That's all you can ask for in any library like this.''} 
\user{3} particularly found the widgets to be `self-explanatory', especially for Angular and Javascript developers.
\user{4} appreciated the consistent design of the widgets and the ability to externally modify the provenance.}

\paragraphHeadingSpace\bpstart{\add{Development Effort.}} In terms of overall development effort, \user{1} took ~7 hours whereas \user{2, 3, 4} took 3-4 hours to set up their visualization system and integrate \app.
While \user{1} found the study to be \emph{``time-consuming''} but \emph{``really fun''}, \user{2, 3, 4} found the amount of time and effort to be appropriate. \user{2} did not find a very steep learning curve and said, \emph{```Intuitive' will be a very good word to describe it.''}

\paragraphHeadingSpace\bpstart{\add{Development Strategies.}} 
Participants found the documentation (\user{1,2,3}) and starter code (\user{2,3}) helpful.
\user{2} said, \emph{``The sample code to start with was just very helpful because I pretty much just modified it to suit my use-case and it just worked. This single-handedly cut the amount of time I spent in half.''}
\add{\user{2,3} requested adding more advanced examples in the eventual documentation.}
\add{\user{1} accessed the original PrimeNG and ngx-slider documentations to try and customize the dropdown options and slider options, respectively. 
\user{2,4} requested alternate widget layouts, e.g., horizontally laid out radio buttons and checkboxes (\user{2}) and vertically laid out sliders (\user{4}).
These customizations and configurations are currently restricted and unsupported, respectively, because they would conflict with the provenance overlay implementation. A takeaway for us is to acknowledge these limitations in the library documentation and include a roadmap for future features.}

\subsubsection{Enhancements and Feature Requests.}
Our developer participants suggested enhancements to \app including \edit{additional visualizations (\user{1, 4}),
custom glyphs for checkboxes (\user{3}),} responsiveness to browser window resize events (\user{1,4}), better tooltip positioning (\user{4}), and \add{better support for styling and theming (\user{2,3})}.
\user{4} suggested incorporating selective tracking for recency and frequency, and high-precision range sliders for more granular and precise filtering in temporal views. \user{4} also suggested enhancements for making the tooltips more human-friendly, such as lowering the precision of numeric attributes (e.g., attack=33.348375), and using 1-based indexing for displaying interactions, instead of `0th/7th interaction'.

\section{Discussion}
\label{section:discussion}
\subsection{Limitations and Future Work}

\add{\app may require developers to understand certain core concepts of its dependencies (PrimeNG, ngx-slider, and Angular), which might lead to issues and limitations, necessitating workarounds.}
\edit{For example, customizing the option templates in the dropdowns and re-orienting the sliders, radio buttons, and checkboxes is currently restricted as it conflicts with the provenance overlays.}
\add{Future work is planned to ensure \app inherits all base library features.}

\add{Next, \app also inherits the inherent limitations of standard UI controls pertaining to scalability and usability. For instance, sliders and dropdowns often struggle with large ranges and numerous options, respectively. 
As a workaround, developers can increase a slider's step size (reducing the number of selectable values), improving usability but sacrificing precision; and dropdown options can be reordered or filtered based on frequency or recency to ensure already interacted options are always visible and accessible.}

\add{Next, visualizations often involve multi-dimensional interactions like brushing and linking~\cite{keim2002information} or smart brushing~\cite{roberts2018smart}, wherein multiple attributes get modified in the same interaction. \app can currently visualize such provenance independently on each widget (as demonstrated in Section~\ref{subsubsection:vistowidget}, Figure~\ref{fig:vis-to-widget}), leading to potential misrepresentation and information loss.
Future work is planned to track and visualize provenance across multiple widgets.}

Lastly, we plan to provide a more flexible and extensible API that will support more widgets, allow customization of existing widgets (e.g., placing provenance scents more freely, e.g., use stroke and opacity instead of color and size), and facilitate creation of new custom widgets (e.g., date pickers). We will also include the ability to compute additional metrics beyond frequency and recency (e.g., their combination).

\subsection{Research Opportunities}
\add{In this work, we evaluated how visualization developers can configure \app, but it remains to be studied how well the widgets can enhance users' visual data analysis workflows. Below we present relevant research opportunities to pursue in the future.}

\paragraphHeadingSpace\bpstart{Systematically Studying Analytical Provenance.}
ProvenanceWidgets can be used to compute advanced metrics to understand co-usage behavior across multiple UI controls, detect biases in control panel usage, and identify over-explored or under-explored aspects of data~\cite{narechania2021lumos}. Such analyses could provide valuable insights into user behavior and guide interface improvements to mitigate biases and optimize user workflows.

\paragraphHeadingSpace\bpstart{Designing Control Panels.}
Oftentimes, certain UI controls receive disproportionate usage due to their visibility or positioning. ProvenanceWidgets could offer solutions such as sorting or filtering affordances to re-layout the control panel widgets based on usage statistics. Additionally, visual cues like usage provenance indicators could be implemented to provide users with insights into the popularity of different controls, promoting more uniform interaction behaviors.

\paragraphHeadingSpace\bpstart{Facilitating Co-Adaptive Guidance.}
\app can be extended to facilitate co-adaptive guidance~\cite{sperrle2021co}, where UI elements dynamically adapt to user interactions in real-time. 
Such systems can provide personalized assistance while optimizing overall analytical workflows, fostering a more symbiotic relationship with the user.

\section{Conclusion}
\label{section:conclusion}
We presented \app, a Javascript library of UI control elements to track and dynamically overlay a user's analytic provenance (i.e., the history of their interactions with the UI control elements).
\edit{Using \app, we recreated three prior widget libraries: (1) Scented Widgets, (2) Phosphor objects, and (3) Dynamic Query Widgets.
Case studies with four developers revealed the effectiveness of \app to build custom provenance-based applications.} 
\edit{ProvenanceWidgets is available as open-source software at \url{https://github.com/ProvenanceWidgets}.}

\clearpage
\section*{Supplemental Materials}
\label{sec:supplemental_materials}


 
\edit{\app is available as open-source software at \url{https://github.com/ProvenanceWidgets/ProvenanceWidgets}.}
\edit{Supplemental material} is available in the IEEE Xplore digital repository as well as Github (\url{https://github.com/ProvenanceWidgets/Supplemental-Material}); it includes (1) a video demonstration of \app, (2) a design document illustrating the alternative designs we explored during the design and development of \app, (3) the application source code demonstrating \app replicating prior work, and (4) the detailed information including source code of developer case studies that we used for evaluating \app.





\acknowledgments{%
This material is based upon work supported by NSF IIS-1750474. We are grateful to our user study participants, members of the Georgia Tech Visualization Lab, and anonymous reviewers for their feedback at different stages of this work.
We used Google Scholar~\cite{googlescholar} and vitaLITy~\cite{vitality2022narechania} to assist us during our literature review.}

\bibliographystyle{abbrv-doi-hyperref}

\bibliography{template}

\begin{thebibliography}{10}

\bibitem{aigner2013evalbench}
W.~Aigner, S.~Hoffmann, and A.~Rind.
\newblock {Evalbench: A software library for visualization evaluation}.
\newblock In {\em {Computer Graphics Forum}}, vol.~32, pp. 41--50. Wiley Online Library, 2013. \href{https://doi.org/10.1111/cgf.12091}
{doi: {{%
10\hspace{.1pt}\discretionary{.}{%
}{.}\hspace{.4pt}1111\discretionary{/}{%
}{/}cgf\hspace{.1pt}\discretionary{.}{%
}{.}\hspace{.4pt}12091}}}


\bibitem{alexander2009revisiting}
J.~Alexander, A.~Cockburn, S.~Fitchett, C.~Gutwin, and S.~Greenberg.
\newblock {Revisiting read wear: analysis, design, and evaluation of a footprints scrollbar}.
\newblock In {\em {Proceedings of the SIGCHI Conference on Human Factors in Computing Systems}}, pp. 1665--1674, 2009. \href{https://doi.org/10.1145/1518701.1518957}
{doi: {{%
10\hspace{.1pt}\discretionary{.}{%
}{.}\hspace{.4pt}1145\discretionary{/}{%
}{/}1518701\hspace{.1pt}\discretionary{.}{%
}{.}\hspace{.4pt}1518957}}}


\bibitem{angular}
{Angular}.
\newblock \url{https://angular.io}.
\newblock Accessed: March 31, 2024.

\bibitem{angularmaterial}
{Angular Material}.
\newblock \url{https://material.angular.io}, 2024.

\bibitem{arroyo2006usability}
E.~Arroyo, T.~Selker, and W.~Wei.
\newblock {Usability tool for analysis of web designs using mouse tracks}.
\newblock In {\em {CHI'06 extended abstracts on Human factors in computing systems}}, pp. 484--489, 2006. \href{https://doi.org/10.1145/1125451.1125557}
{doi: {{%
10\hspace{.1pt}\discretionary{.}{%
}{.}\hspace{.4pt}1145\discretionary{/}{%
}{/}1125451\hspace{.1pt}\discretionary{.}{%
}{.}\hspace{.4pt}1125557}}}


\bibitem{badam2017supporting}
S.~K. Badam, Z.~Zeng, E.~Wall, A.~Endert, and N.~Elmqvist.
\newblock {Supporting Team-First Visual Analytics through Group Activity Representations.}
\newblock In {\em {Graphics Interface}}, pp. 208--213, 2017. \href{https://doi.org/10.5555/3141475.3141515}
{doi: {{%
10\hspace{.1pt}\discretionary{.}{%
}{.}\hspace{.4pt}5555\discretionary{/}{%
}{/}3141475\hspace{.1pt}\discretionary{.}{%
}{.}\hspace{.4pt}3141515}}}


\bibitem{baudisch2006phosphor}
P.~Baudisch, D.~Tan, M.~Collomb, D.~Robbins, K.~Hinckley, M.~Agrawala, S.~Zhao, and G.~Ramos.
\newblock {Phosphor: explaining transitions in the user interface using afterglow effects}.
\newblock In {\em {Proceedings of the 19th annual ACM symposium on User interface software and technology}}, pp. 169--178, 2006. \href{https://doi.org/10.1145/1166253.1166280}
{doi: {{%
10\hspace{.1pt}\discretionary{.}{%
}{.}\hspace{.4pt}1145\discretionary{/}{%
}{/}1166253\hspace{.1pt}\discretionary{.}{%
}{.}\hspace{.4pt}1166280}}}


\bibitem{bavoil2005vistrails}
L.~Bavoil, S.~P. Callahan, P.~J. Crossno, J.~Freire, C.~E. Scheidegger, C.~T. Silva, and H.~T. Vo.
\newblock {Vistrails: Enabling interactive multiple-view visualizations}.
\newblock In {\em {VIS 05. IEEE Visualization, 2005.}}, pp. 135--142. IEEE, 2005. \href{https://doi.org/10.1109/VISUAL.2005.1532788}
{doi: {{%
10\hspace{.1pt}\discretionary{.}{%
}{.}\hspace{.4pt}1109\discretionary{/}{%
}{/}VISUAL\hspace{.1pt}\discretionary{.}{%
}{.}\hspace{.4pt}2005\hspace{.1pt}\discretionary{.}{%
}{.}\hspace{.4pt}1532788}}}


\bibitem{blackwell2001cognitive}
A.~F. Blackwell, C.~Britton, A.~Cox, T.~R. Green, C.~Gurr, G.~Kadoda, M.~S. Kutar, M.~Loomes, C.~L. Nehaniv, M.~Petre, et~al.
\newblock {Cognitive dimensions of notations: Design tools for cognitive technology}.
\newblock In {\em {Cognitive Technology: Instruments of Mind: 4th International Conference, CT 2001 Coventry, UK, August 6--9, 2001 Proceedings}}, pp. 325--341. Springer, 2001. \href{https://doi.org/10.1007/3-540-44617-6_31}
{doi: {{%
10\hspace{.1pt}\discretionary{.}{%
}{.}\hspace{.4pt}1007\discretionary{/}{%
}{/}3\discretionary{%
}{-}{-}540\discretionary{%
}{-}{-}44617\discretionary{%
}{-}{-}6\_31}}}


\bibitem{block2022influence}
J.~E. Block, S.~Esmaeili, E.~D. Ragan, J.~R. Goodall, and G.~D. Richardson.
\newblock {The Influence of Visual Provenance Representations on Strategies in a Collaborative Hand-off Data Analysis Scenario}.
\newblock {\em {IEEE Transactions on Visualization and Computer Graphics}}, 29(1):1113--1123, 2023. \href{https://doi.org/10.1109/TVCG.2022.3209495}
{doi: {{%
10\hspace{.1pt}\discretionary{.}{%
}{.}\hspace{.4pt}1109\discretionary{/}{%
}{/}TVCG\hspace{.1pt}\discretionary{.}{%
}{.}\hspace{.4pt}2022\hspace{.1pt}\discretionary{.}{%
}{.}\hspace{.4pt}3209495}}}


\bibitem{bootstrap}
{Bootstrap}.
\newblock \url{https://getbootstrap.com}, 2024.

\bibitem{boyatzis1998transforming}
R.~E. Boyatzis.
\newblock {\em {Transforming Qualitative Information: Thematic Analysis and Code Development.}}
\newblock {Sage Publications}, 1998.

\bibitem{callahan2006vistrails}
S.~P. Callahan, J.~Freire, E.~Santos, C.~E. Scheidegger, C.~T. Silva, and H.~T. Vo.
\newblock {VisTrails: visualization meets data management}.
\newblock In {\em {Proceedings of the 2006 ACM SIGMOD international conference on Management of data}}, pp. 745--747, 2006. \href{https://doi.org/10.1145/1142473.1142574}
{doi: {{%
10\hspace{.1pt}\discretionary{.}{%
}{.}\hspace{.4pt}1145\discretionary{/}{%
}{/}1142473\hspace{.1pt}\discretionary{.}{%
}{.}\hspace{.4pt}1142574}}}


\bibitem{cernea2013emotion}
D.~Cernea, C.~Weber, A.~Ebert, and A.~Kerren.
\newblock {Emotion scents: a method of representing user emotions on gui widgets}.
\newblock In {\em {Visualization and Data Analysis 2013}}, vol. 8654, pp. 168--181. SPIE, 2013. \href{https://doi.org/10.1117/12.2001261}
{doi: {{%
10\hspace{.1pt}\discretionary{.}{%
}{.}\hspace{.4pt}1117\discretionary{/}{%
}{/}12\hspace{.1pt}\discretionary{.}{%
}{.}\hspace{.4pt}2001261}}}


\bibitem{chung2015guidelines}
G.~K. Chung.
\newblock {Guidelines for the design and implementation of game telemetry for serious games analytics}.
\newblock {\em {Serious games analytics: Methodologies for performance measurement, assessment, and improvement}}, pp. 59--79, 2015. \href{https://doi.org/10.1007/978-3-319-05834-4_3}
{doi: {{%
10\hspace{.1pt}\discretionary{.}{%
}{.}\hspace{.4pt}1007\discretionary{/}{%
}{/}978\discretionary{%
}{-}{-}3\discretionary{%
}{-}{-}319\discretionary{%
}{-}{-}05834\discretionary{%
}{-}{-}4\_3}}}


\bibitem{cutler2020trrack}
Z.~Cutler, K.~Gadhave, and A.~Lex.
\newblock {Trrack: A library for provenance-tracking in web-based visualizations}.
\newblock In {\em {2020 IEEE Visualization Conference (VIS)}}, pp. 116--120. IEEE, 2020. \href{https://doi.org/10.1109/VIS47514.2020.00030}
{doi: {{%
10\hspace{.1pt}\discretionary{.}{%
}{.}\hspace{.4pt}1109\discretionary{/}{%
}{/}VIS47514\hspace{.1pt}\discretionary{.}{%
}{.}\hspace{.4pt}2020\hspace{.1pt}\discretionary{.}{%
}{.}\hspace{.4pt}00030}}}


\bibitem{revisit2023ding}
Y.~Ding, J.~Wilburn, H.~Shrestha, A.~Ndlovu, K.~Gadhave, C.~Nobre, A.~Lex, and L.~Harrison.
\newblock {reVISit: Supporting Scalable Evaluation of Interactive Visualizations}.
\newblock In {\em {2023 IEEE Visualization and Visual Analytics (VIS)}}, pp. 31--35, 2023. \href{https://doi.org/10.1109/VIS54172.2023.00015}
{doi: {{%
10\hspace{.1pt}\discretionary{.}{%
}{.}\hspace{.4pt}1109\discretionary{/}{%
}{/}VIS54172\hspace{.1pt}\discretionary{.}{%
}{.}\hspace{.4pt}2023\hspace{.1pt}\discretionary{.}{%
}{.}\hspace{.4pt}00015}}}


\bibitem{drachen2015behavioral}
A.~Drachen.
\newblock {Behavioral telemetry in games user research}.
\newblock {\em {Game user experience evaluation}}, pp. 135--165, 2015. \href{https://doi.org/10.1007/978-3-319-15985-0_7}
{doi: {{%
10\hspace{.1pt}\discretionary{.}{%
}{.}\hspace{.4pt}1007\discretionary{/}{%
}{/}978\discretionary{%
}{-}{-}3\discretionary{%
}{-}{-}319\discretionary{%
}{-}{-}15985\discretionary{%
}{-}{-}0\_7}}}


\bibitem{drachen2013game}
A.~Drachen, M.~Seif El-Nasr, and A.~Canossa.
\newblock {Game analytics--the basics}.
\newblock {\em {Game analytics: Maximizing the value of player data}}, pp. 13--40, 2013. \href{https://doi.org/10.1007/978-1-4471-4769-5}
{doi: {{%
10\hspace{.1pt}\discretionary{.}{%
}{.}\hspace{.4pt}1007\discretionary{/}{%
}{/}978\discretionary{%
}{-}{-}1\discretionary{%
}{-}{-}4471\discretionary{%
}{-}{-}4769\discretionary{%
}{-}{-}5}}}


\bibitem{drawio}
{Draw.io}.
\newblock \url{https://www.draw.io}.
\newblock Accessed: March 31, 2024.

\bibitem{dunne2012graphtrail}
C.~Dunne, N.~Henry~Riche, B.~Lee, R.~Metoyer, and G.~Robertson.
\newblock {GraphTrail: Analyzing large multivariate, heterogeneous networks while supporting exploration history}.
\newblock In {\em {Proceedings of the SIGCHI Conference on Human Factors in Computing Systems}}, pp. 1663--1672, 2012. \href{https://doi.org/10.1145/2207676.2208293}
{doi: {{%
10\hspace{.1pt}\discretionary{.}{%
}{.}\hspace{.4pt}1145\discretionary{/}{%
}{/}2207676\hspace{.1pt}\discretionary{.}{%
}{.}\hspace{.4pt}2208293}}}


\bibitem{2024_loops}
K.~Eckelt, K.~Gadhave, A.~Lex, and M.~Streit.
\newblock {Loops: Leveraging Provenance and Visualization to Support Exploratory Data Analysis in Notebooks}.
\newblock {\em {OSF Preprint}}, 2023. \href{https://doi.org/10.31219/osf.io/79eyn}
{doi: {{%
10\hspace{.1pt}\discretionary{.}{%
}{.}\hspace{.4pt}31219\discretionary{/}{%
}{/}osf\hspace{.1pt}\discretionary{.}{%
}{.}\hspace{.4pt}io\discretionary{/}{%
}{/}79eyn}}}


\bibitem{epperson2022leveraging}
W.~Epperson, D.~Jung-Lin~Lee, L.~Wang, K.~Agarwal, A.~G. Parameswaran, D.~Moritz, and A.~Perer.
\newblock {Leveraging analysis history for improved in situ visualization recommendation}.
\newblock In {\em {Computer Graphics Forum}}, vol.~41, pp. 145--155. Wiley Online Library, 2022. \href{https://doi.org/10.1111/cgf.14529}
{doi: {{%
10\hspace{.1pt}\discretionary{.}{%
}{.}\hspace{.4pt}1111\discretionary{/}{%
}{/}cgf\hspace{.1pt}\discretionary{.}{%
}{.}\hspace{.4pt}14529}}}


\bibitem{feng2016hindsight}
M.~Feng, C.~Deng, E.~M. Peck, and L.~Harrison.
\newblock {Hindsight: Encouraging exploration through direct encoding of personal interaction history}.
\newblock {\em {IEEE Transactions on Visualization and Computer Graphics}}, 23(1):351--360, 2017. \href{https://doi.org/10.1109/TVCG.2016.2599058}
{doi: {{%
10\hspace{.1pt}\discretionary{.}{%
}{.}\hspace{.4pt}1109\discretionary{/}{%
}{/}TVCG\hspace{.1pt}\discretionary{.}{%
}{.}\hspace{.4pt}2016\hspace{.1pt}\discretionary{.}{%
}{.}\hspace{.4pt}2599058}}}


\bibitem{fullstory}
{FullStory}.
\newblock \url{https://www.fullstory.com}.
\newblock Accessed: June 15, 2024.

\bibitem{gagne2011deeper}
A.~R. Gagn{\'e}, M.~S. El-Nasr, and C.~D. Shaw.
\newblock A deeper look at the use of telemetry for analysis of player behavior in rts games.
\newblock In {\em {International Conference on Entertainment Computing}}, pp. 247--257. Springer, 2011. \href{https://doi.org/10.1007/978-3-642-24500-8_26}
{doi: {{%
10\hspace{.1pt}\discretionary{.}{%
}{.}\hspace{.4pt}1007\discretionary{/}{%
}{/}978\discretionary{%
}{-}{-}3\discretionary{%
}{-}{-}642\discretionary{%
}{-}{-}24500\discretionary{%
}{-}{-}8\_26}}}


\bibitem{googleanalytics}
{Google Analytics}.
\newblock \url{https://marketingplatform.google.com/about/analytics}.
\newblock Accessed: June 15, 2024.

\bibitem{googlescholar}
{Google Scholar}.
\newblock \url{https://scholar.google.com}, 2024.

\bibitem{gutwin2002traces}
C.~Gutwin.
\newblock {Traces: Visualizing the immediate past to support group interaction}.
\newblock In {\em {Graphics interface}}, pp. 43--50. Citeseer, 2002. \href{https://doi.org/10.20380/GI2002.06}
{doi: {{%
10\hspace{.1pt}\discretionary{.}{%
}{.}\hspace{.4pt}20380\discretionary{/}{%
}{/}GI2002\hspace{.1pt}\discretionary{.}{%
}{.}\hspace{.4pt}06}}}


\bibitem{heer2008generalized}
J.~Heer, M.~Agrawala, and W.~Willett.
\newblock {Generalized selection via interactive query relaxation}.
\newblock In {\em {Proceedings of the SIGCHI Conference on Human Factors in Computing Systems}}, pp. 959--968, 2008. \href{https://doi.org/10.1145/1357054.1357203}
{doi: {{%
10\hspace{.1pt}\discretionary{.}{%
}{.}\hspace{.4pt}1145\discretionary{/}{%
}{/}1357054\hspace{.1pt}\discretionary{.}{%
}{.}\hspace{.4pt}1357203}}}


\bibitem{heer2008graphical}
J.~Heer, J.~Mackinlay, C.~Stolte, and M.~Agrawala.
\newblock {Graphical histories for visualization: Supporting analysis, communication, and evaluation}.
\newblock {\em {IEEE Transactions on Visualization and Computer Graphics}}, 14(6):1189--1196, 2008. \href{https://doi.org/10.1109/TVCG.2008.137}
{doi: {{%
10\hspace{.1pt}\discretionary{.}{%
}{.}\hspace{.4pt}1109\discretionary{/}{%
}{/}TVCG\hspace{.1pt}\discretionary{.}{%
}{.}\hspace{.4pt}2008\hspace{.1pt}\discretionary{.}{%
}{.}\hspace{.4pt}137}}}


\bibitem{hill2003awareness}
J.~Hill and C.~Gutwin.
\newblock {Awareness support in a groupware widget toolkit}.
\newblock In {\em {Proceedings of the 2003 international ACM SIGGROUP conference on Supporting group work}}, pp. 258--267, 2003. \href{https://doi.org/10.1145/958160.958201}
{doi: {{%
10\hspace{.1pt}\discretionary{.}{%
}{.}\hspace{.4pt}1145\discretionary{/}{%
}{/}958160\hspace{.1pt}\discretionary{.}{%
}{.}\hspace{.4pt}958201}}}


\bibitem{hill1992edit}
W.~C. Hill, J.~D. Hollan, D.~Wroblewski, and T.~McCandless.
\newblock {Edit wear and read wear}.
\newblock In {\em {Proceedings of the SIGCHI Conference on Human Factors in Computing Systems}}, pp. 3--9, 1992. \href{https://doi.org/10.1145/142750.142751}
{doi: {{%
10\hspace{.1pt}\discretionary{.}{%
}{.}\hspace{.4pt}1145\discretionary{/}{%
}{/}142750\hspace{.1pt}\discretionary{.}{%
}{.}\hspace{.4pt}142751}}}


\bibitem{hotjar}
{Hotjar}.
\newblock \url{https://www.hotjar.com}.
\newblock Accessed: June 15, 2024.

\bibitem{jacob2017oh}
L.~Jacob, E.~Clua, and D.~de~Oliveira.
\newblock {Oh Gosh!! Why is this game so hard? Identifying cycle patterns in 2D platform games using provenance data}.
\newblock {\em {Entertainment Computing}}, 19:65--81, 2017. \href{https://doi.org/10.1016/j.entcom.2016.12.002}
{doi: {{%
10\hspace{.1pt}\discretionary{.}{%
}{.}\hspace{.4pt}1016\discretionary{/}{%
}{/}j\hspace{.1pt}\discretionary{.}{%
}{.}\hspace{.4pt}entcom\hspace{.1pt}\discretionary{.}{%
}{.}\hspace{.4pt}2016\hspace{.1pt}\discretionary{.}{%
}{.}\hspace{.4pt}12\hspace{.1pt}\discretionary{.}{%
}{.}\hspace{.4pt}002}}}


\bibitem{jqueryui}
{jQuery UI}.
\newblock \url{https://jqueryui.com}, 2024.

\bibitem{keim2002information}
D.~A. Keim.
\newblock {Information visualization and visual data mining}.
\newblock {\em {IEEE Transactions on Visualization and Computer Graphics}}, 8(1):1--8, 2002. \href{https://doi.org/10.1109/2945.981847}
{doi: {{%
10\hspace{.1pt}\discretionary{.}{%
}{.}\hspace{.4pt}1109\discretionary{/}{%
}{/}2945\hspace{.1pt}\discretionary{.}{%
}{.}\hspace{.4pt}981847}}}


\bibitem{kohwalter2018understanding}
T.~C. Kohwalter, F.~M. de~Azeredo~Figueira, E.~A. de~Lima~Serdeiro, J.~R. da~Silva~Junior, L.~G.~P. Murta, and E.~W.~G. Clua.
\newblock {Understanding game sessions through provenance}.
\newblock {\em {Entertainment Computing}}, 27:110--127, 2018. \href{https://doi.org/10.1016/j.entcom.2018.05.001}
{doi: {{%
10\hspace{.1pt}\discretionary{.}{%
}{.}\hspace{.4pt}1016\discretionary{/}{%
}{/}j\hspace{.1pt}\discretionary{.}{%
}{.}\hspace{.4pt}entcom\hspace{.1pt}\discretionary{.}{%
}{.}\hspace{.4pt}2018\hspace{.1pt}\discretionary{.}{%
}{.}\hspace{.4pt}05\hspace{.1pt}\discretionary{.}{%
}{.}\hspace{.4pt}001}}}


\bibitem{kohwalter2020provchastic}
T.~C. Kohwalter, L.~G. Murta, and E.~W. Clua.
\newblock {Provchastic: Understanding and predicting game events using provenance}.
\newblock In {\em {International Conference on Entertainment Computing}}, pp. 90--103. Springer, 2020. \href{https://doi.org/10.1007/978-3-030-65736-9_7}
{doi: {{%
10\hspace{.1pt}\discretionary{.}{%
}{.}\hspace{.4pt}1007\discretionary{/}{%
}{/}978\discretionary{%
}{-}{-}3\discretionary{%
}{-}{-}030\discretionary{%
}{-}{-}65736\discretionary{%
}{-}{-}9\_7}}}


\bibitem{kohwalter2017capturing}
T.~C. Kohwalter, L.~G.~P. Murta, and E.~W.~G. Clua.
\newblock {Capturing game telemetry with provenance}.
\newblock In {\em {2017 16th Brazilian Symposium on Computer Games and Digital Entertainment (SBGames)}}, pp. 66--75. IEEE, 2017. \href{https://doi.org/10.1109/SBGames.2017.00016}
{doi: {{%
10\hspace{.1pt}\discretionary{.}{%
}{.}\hspace{.4pt}1109\discretionary{/}{%
}{/}SBGames\hspace{.1pt}\discretionary{.}{%
}{.}\hspace{.4pt}2017\hspace{.1pt}\discretionary{.}{%
}{.}\hspace{.4pt}00016}}}


\bibitem{lim2015toward}
C.-U. Lim and D.~F. Harrell.
\newblock {Toward telemetry-driven analytics for understanding players and their avatars in videogames}.
\newblock In {\em {Proceedings of the 33rd Annual ACM Conference Extended Abstracts on Human Factors in Computing Systems}}, pp. 1175--1180, 2015. \href{https://doi.org/10.1145/2702613.2732783}
{doi: {{%
10\hspace{.1pt}\discretionary{.}{%
}{.}\hspace{.4pt}1145\discretionary{/}{%
}{/}2702613\hspace{.1pt}\discretionary{.}{%
}{.}\hspace{.4pt}2732783}}}


\bibitem{liu2014effects}
Z.~Liu and J.~Heer.
\newblock {The effects of interactive latency on exploratory visual analysis}.
\newblock {\em {IEEE Transactions on Visualization and Computer Graphics}}, 20(12):2122--2131, 2014. \href{https://doi.org/10.1109/TVCG.2014.2346452}
{doi: {{%
10\hspace{.1pt}\discretionary{.}{%
}{.}\hspace{.4pt}1109\discretionary{/}{%
}{/}TVCG\hspace{.1pt}\discretionary{.}{%
}{.}\hspace{.4pt}2014\hspace{.1pt}\discretionary{.}{%
}{.}\hspace{.4pt}2346452}}}


\bibitem{madanagopal2019analytic}
K.~Madanagopal, E.~D. Ragan, and P.~Benjamin.
\newblock {Analytic provenance in practice: The role of provenance in real-world visualization and data analysis environments}.
\newblock {\em {IEEE Computer Graphics and Applications}}, 39(6):30--45, 2019. \href{https://doi.org/10.1109/MCG.2019.2933419}
{doi: {{%
10\hspace{.1pt}\discretionary{.}{%
}{.}\hspace{.4pt}1109\discretionary{/}{%
}{/}MCG\hspace{.1pt}\discretionary{.}{%
}{.}\hspace{.4pt}2019\hspace{.1pt}\discretionary{.}{%
}{.}\hspace{.4pt}2933419}}}


\bibitem{materialui}
{Material-UI}.
\newblock \url{https://material-ui.com}, 2024.

\bibitem{mcinnes2018umap}
L.~McInnes, J.~Healy, and J.~Melville.
\newblock {Umap: Uniform manifold approximation and projection for dimension reduction}.
\newblock {\em {arXiv preprint arXiv:1802.03426}}, 2018. \href{https://doi.org/10.21105/joss.00861}
{doi: {{%
10\hspace{.1pt}\discretionary{.}{%
}{.}\hspace{.4pt}21105\discretionary{/}{%
}{/}joss\hspace{.1pt}\discretionary{.}{%
}{.}\hspace{.4pt}00861}}}


\bibitem{melo2020player}
S.~A. Melo, T.~C. Kohwalter, E.~Clua, A.~Paes, and L.~Murta.
\newblock {Player behavior profiling through provenance graphs and representation learning}.
\newblock In {\em {Proceedings of the 15th International Conference on the Foundations of Digital Games}}, pp. 1--11, 2020. \href{https://doi.org/10.1145/3402942.3402961}
{doi: {{%
10\hspace{.1pt}\discretionary{.}{%
}{.}\hspace{.4pt}1145\discretionary{/}{%
}{/}3402942\hspace{.1pt}\discretionary{.}{%
}{.}\hspace{.4pt}3402961}}}


\bibitem{miller1994magical}
G.~A. Miller.
\newblock {The magical number seven, plus or minus two: Some limits on our capacity for processing information.}
\newblock {\em {Psychological review}}, 101(2):343, 1994. \href{https://doi.org/10.1037/h0043158}
{doi: {{%
10\hspace{.1pt}\discretionary{.}{%
}{.}\hspace{.4pt}1037\discretionary{/}{%
}{/}h0043158}}}


\bibitem{mixpanel}
{Mixpanel}.
\newblock \url{https://mixpanel.com}.
\newblock Accessed: June 15, 2024.

\bibitem{mouseflow}
{Mouseflow}.
\newblock \url{https://mouseflow.com}.
\newblock Accessed: June 15, 2024.

\bibitem{narechania2021lumos}
A.~Narechania, A.~Coscia, E.~Wall, and A.~Endert.
\newblock {Lumos: Increasing Awareness of Analytic Behavior during Visual Data Analysis}.
\newblock {\em {IEEE Transactions on Visualization and Computer Graphics}}, 28(1):1009--1018, 2022. \href{https://doi.org/10.1109/TVCG.2021.3114827}
{doi: {{%
10\hspace{.1pt}\discretionary{.}{%
}{.}\hspace{.4pt}1109\discretionary{/}{%
}{/}TVCG\hspace{.1pt}\discretionary{.}{%
}{.}\hspace{.4pt}2021\hspace{.1pt}\discretionary{.}{%
}{.}\hspace{.4pt}3114827}}}


\bibitem{vitality2022narechania}
A.~Narechania, A.~Karduni, R.~Wesslen, and E.~Wall.
\newblock {vitaLITy: Promoting Serendipitous Discovery of Academic Literature with Transformers \& Visual Analytics}.
\newblock {\em IEEE Transactions on Visualization and Computer Graphics}, 28(1):486--496, 2022. \href{https://doi.org/10.1109/TVCG.2021.3114820}
{doi: {{%
10\hspace{.1pt}\discretionary{.}{%
}{.}\hspace{.4pt}1109\discretionary{/}{%
}{/}TVCG\hspace{.1pt}\discretionary{.}{%
}{.}\hspace{.4pt}2021\hspace{.1pt}\discretionary{.}{%
}{.}\hspace{.4pt}3114820}}}


\bibitem{newrelic}
{New Relic}.
\newblock \url{https://newrelic.com}.
\newblock Accessed: June 15, 2024.

\bibitem{nielsen2010eyetracking}
J.~Nielsen and K.~Pernice.
\newblock {\em {Eyetracking web usability}}.
\newblock New Riders, 2010.

\bibitem{nobre2021revisit}
C.~Nobre, D.~Wootton, Z.~Cutler, L.~Harrison, H.~Pfister, and A.~Lex.
\newblock {reVISit: Looking under the hood of interactive visualization studies}.
\newblock In {\em {Proceedings of the 2021 CHI Conference on Human Factors in Computing Systems}}, pp. 1--13, 2021. \href{https://doi.org/10.1145/3411764.3445382}
{doi: {{%
10\hspace{.1pt}\discretionary{.}{%
}{.}\hspace{.4pt}1145\discretionary{/}{%
}{/}3411764\hspace{.1pt}\discretionary{.}{%
}{.}\hspace{.4pt}3445382}}}


\bibitem{north2011analytic}
C.~North, R.~Chang, A.~Endert, W.~Dou, R.~May, B.~Pike, and G.~Fink.
\newblock {Analytic provenance: process+ interaction+ insight}.
\newblock In {\em {CHI'11 Extended Abstracts on Human Factors in Computing Systems}}, pp. 33--36. 2011. \href{https://doi.org/10.1145/1979742.1979570}
{doi: {{%
10\hspace{.1pt}\discretionary{.}{%
}{.}\hspace{.4pt}1145\discretionary{/}{%
}{/}1979742\hspace{.1pt}\discretionary{.}{%
}{.}\hspace{.4pt}1979570}}}


\bibitem{okoe2015graphunit}
M.~Okoe and R.~Jianu.
\newblock {Graphunit: Evaluating interactive graph visualizations using crowdsourcing}.
\newblock In {\em {Computer Graphics Forum}}, vol.~34, pp. 451--460. Wiley Online Library, 2015. \href{https://doi.org/10.1111/cgf.12657}
{doi: {{%
10\hspace{.1pt}\discretionary{.}{%
}{.}\hspace{.4pt}1111\discretionary{/}{%
}{/}cgf\hspace{.1pt}\discretionary{.}{%
}{.}\hspace{.4pt}12657}}}


\bibitem{paden2024biasbuzz}
J.~R. Paden, A.~Narechania, and A.~Endert.
\newblock {BiasBuzz: Combining Visual Guidance with Haptic Feedback to Increase Awareness of Analytic Behavior during Visual Data Analysis}.
\newblock In {\em {Extended Abstracts of the CHI Conference on Human Factors in Computing Systems}}, pp. 1--7, 2024. \href{https://doi.org/10.1145/3613905.3651064}
{doi: {{%
10\hspace{.1pt}\discretionary{.}{%
}{.}\hspace{.4pt}1145\discretionary{/}{%
}{/}3613905\hspace{.1pt}\discretionary{.}{%
}{.}\hspace{.4pt}3651064}}}


\bibitem{primeng}
{PrimeNG}.
\newblock \url{https://www.primefaces.org/primeng}, 2024.

\bibitem{ragan2015characterizing}
E.~D. Ragan, A.~Endert, J.~Sanyal, and J.~Chen.
\newblock {Characterizing provenance in visualization and data analysis: an organizational framework of provenance types and purposes}.
\newblock {\em {IEEE Transactions on Visualization and Computer Graphics}}, 22(1):31--40, 2016. \href{https://doi.org/10.1109/TVCG.2015.2467551}
{doi: {{%
10\hspace{.1pt}\discretionary{.}{%
}{.}\hspace{.4pt}1109\discretionary{/}{%
}{/}TVCG\hspace{.1pt}\discretionary{.}{%
}{.}\hspace{.4pt}2015\hspace{.1pt}\discretionary{.}{%
}{.}\hspace{.4pt}2467551}}}


\bibitem{react}
{React}.
\newblock \url{https://reactjs.org}.
\newblock Accessed: March 31, 2024.

\bibitem{reactbootstrap}
{React Bootstrap}.
\newblock \url{https://react-bootstrap.github.io}, 2024.

\bibitem{roberts2018smart}
R.~C. Roberts, R.~S. Laramee, G.~A. Smith, P.~Brookes, and T.~D'Cruze.
\newblock {Smart brushing for parallel coordinates}.
\newblock {\em {IEEE Transactions on Visualization and Computer Graphics}}, 25(3):1575--1590, 2019. \href{https://doi.org/10.1109/TVCG.2018.2808969}
{doi: {{%
10\hspace{.1pt}\discretionary{.}{%
}{.}\hspace{.4pt}1109\discretionary{/}{%
}{/}TVCG\hspace{.1pt}\discretionary{.}{%
}{.}\hspace{.4pt}2018\hspace{.1pt}\discretionary{.}{%
}{.}\hspace{.4pt}2808969}}}


\bibitem{sarvghad2015exploiting}
A.~Sarvghad and M.~Tory.
\newblock {Exploiting analysis history to support collaborative data analysis}.
\newblock In {\em {Proceedings of the 41st Graphics Interface Conference}}, pp. 123--130, 2015.

\bibitem{satyanarayan2016vega}
A.~Satyanarayan, D.~Moritz, K.~Wongsuphasawat, and J.~Heer.
\newblock {Vega-lite: A grammar of interactive graphics}.
\newblock {\em {IEEE Transactions on Visualization and Computer Graphics}}, 23(1):341--350, 2017. \href{https://doi.org/10.1109/TVCG.2016.2599030}
{doi: {{%
10\hspace{.1pt}\discretionary{.}{%
}{.}\hspace{.4pt}1109\discretionary{/}{%
}{/}TVCG\hspace{.1pt}\discretionary{.}{%
}{.}\hspace{.4pt}2016\hspace{.1pt}\discretionary{.}{%
}{.}\hspace{.4pt}2599030}}}


\bibitem{serebii}
{Serebii.net}.
\newblock \url{http://serebii.net}.
\newblock Accessed: March 31, 2024.

\bibitem{shneiderman1994dynamic}
B.~Shneiderman.
\newblock {Dynamic queries for visual information seeking}.
\newblock {\em {IEEE software}}, 11(6):70--77, 1994. \href{https://doi.org/10.1109/52.329404}
{doi: {{%
10\hspace{.1pt}\discretionary{.}{%
}{.}\hspace{.4pt}1109\discretionary{/}{%
}{/}52\hspace{.1pt}\discretionary{.}{%
}{.}\hspace{.4pt}329404}}}


\bibitem{shneiderman2003eyes}
B.~Shneiderman.
\newblock {The eyes have it: A task by data type taxonomy for information visualizations}.
\newblock In {\em {The craft of information visualization}}, pp. 364--371. Elsevier, 2003. \href{https://doi.org/10.1109/VL.1996.545307}
{doi: {{%
10\hspace{.1pt}\discretionary{.}{%
}{.}\hspace{.4pt}1109\discretionary{/}{%
}{/}VL\hspace{.1pt}\discretionary{.}{%
}{.}\hspace{.4pt}1996\hspace{.1pt}\discretionary{.}{%
}{.}\hspace{.4pt}545307}}}


\bibitem{silva2011using}
C.~T. Silva, E.~Anderson, E.~Santos, and J.~Freire.
\newblock {Using vistrails and provenance for teaching scientific visualization}.
\newblock In {\em {Computer Graphics Forum}}, vol.~30, pp. 75--84. Wiley Online Library, 2011. \href{https://doi.org/10.1111/j.1467-8659.2010.01830.x}
{doi: {{%
10\hspace{.1pt}\discretionary{.}{%
}{.}\hspace{.4pt}1111\discretionary{/}{%
}{/}j\hspace{.1pt}\discretionary{.}{%
}{.}\hspace{.4pt}1467\discretionary{%
}{-}{-}8659\hspace{.1pt}\discretionary{.}{%
}{.}\hspace{.4pt}2010\hspace{.1pt}\discretionary{.}{%
}{.}\hspace{.4pt}01830\hspace{.1pt}\discretionary{.}{%
}{.}\hspace{.4pt}x}}}


\bibitem{skopik2005improving}
A.~Skopik and C.~Gutwin.
\newblock {Improving revisitation in fisheye views with visit wear}.
\newblock In {\em {Proceedings of the SIGCHI Conference on Human Factors in Computing Systems}}, pp. 771--780, 2005. \href{https://doi.org/10.1145/1054972.1055079}
{doi: {{%
10\hspace{.1pt}\discretionary{.}{%
}{.}\hspace{.4pt}1145\discretionary{/}{%
}{/}1054972\hspace{.1pt}\discretionary{.}{%
}{.}\hspace{.4pt}1055079}}}


\bibitem{sperrle2021co}
F.~Sperrle, A.~Jeitler, J.~Bernard, D.~Keim, and M.~El-Assady.
\newblock {Co-adaptive visual data analysis and guidance processes}.
\newblock {\em {Computers \& Graphics}}, 100:93--105, 2021. \href{https://doi.org/10.1016/j.cag.2021.06.016}
{doi: {{%
10\hspace{.1pt}\discretionary{.}{%
}{.}\hspace{.4pt}1016\discretionary{/}{%
}{/}j\hspace{.1pt}\discretionary{.}{%
}{.}\hspace{.4pt}cag\hspace{.1pt}\discretionary{.}{%
}{.}\hspace{.4pt}2021\hspace{.1pt}\discretionary{.}{%
}{.}\hspace{.4pt}06\hspace{.1pt}\discretionary{.}{%
}{.}\hspace{.4pt}016}}}


\bibitem{strauss1998basics}
A.~Strauss and J.~Corbin.
\newblock {\em {Basics of Qualitative Research: Techniques and Procedures for Developing Grounded Theory}}.
\newblock {Sage Publications}, 1998. \href{https://doi.org/10.4135/9781452230153}
{doi: {{%
10\hspace{.1pt}\discretionary{.}{%
}{.}\hspace{.4pt}4135\discretionary{/}{%
}{/}9781452230153}}}


\bibitem{sveltekitui}
{SvelteKit UI}.
\newblock \url{https://kit.svelte.dev}, 2024.

\bibitem{vaithilingam2024dynavis}
P.~Vaithilingam, E.~L. Glassman, J.~P. Inala, and C.~Wang.
\newblock {DynaVis: Dynamically Synthesized UI Widgets for Visualization Editing}.
\newblock In {\em {Proceedings of the CHI Conference on Human Factors in Computing Systems}}, CHI '24,  article no. 985,  17 pages. Association for Computing Machinery, New York, NY, USA, 2024. \href{https://doi.org/10.1145/3613904.3642639}
{doi: {{%
10\hspace{.1pt}\discretionary{.}{%
}{.}\hspace{.4pt}1145\discretionary{/}{%
}{/}3613904\hspace{.1pt}\discretionary{.}{%
}{.}\hspace{.4pt}3642639}}}


\bibitem{vue}
{Vue.js}.
\newblock \url{https://vuejs.org}.
\newblock Accessed: March 31, 2024.

\bibitem{vuetify}
{Vuetify}.
\newblock \url{https://vuetifyjs.com}, 2024.

\bibitem{wall2021lrg}
E.~Wall, A.~Narechania, A.~Coscia, J.~Paden, and A.~Endert.
\newblock {Left, Right, and Gender: Exploring Interaction Traces to Mitigate Human Biases}.
\newblock {\em {IEEE Transactions on Visualization and Computer Graphics}}, 28(1):966--975, 2022. \href{https://doi.org/10.1109/TVCG.2021.3114862}
{doi: {{%
10\hspace{.1pt}\discretionary{.}{%
}{.}\hspace{.4pt}1109\discretionary{/}{%
}{/}TVCG\hspace{.1pt}\discretionary{.}{%
}{.}\hspace{.4pt}2021\hspace{.1pt}\discretionary{.}{%
}{.}\hspace{.4pt}3114862}}}


\bibitem{footstepsvscode}
A.~Wattenberger.
\newblock {Footsteps for VS Code}.
\newblock https://marketplace.visualstudio.com/items?itemName=Wattenberger.footsteps, 2021.

\bibitem{willett2007scented}
W.~Willett, J.~Heer, and M.~Agrawala.
\newblock {Scented Widgets: Improving navigation cues with embedded visualizations}.
\newblock {\em {IEEE Transactions on Visualization and Computer Graphics}}, 13(6):1129--1136, 2007. \href{https://doi.org/10.1109/TVCG.2007.70589}
{doi: {{%
10\hspace{.1pt}\discretionary{.}{%
}{.}\hspace{.4pt}1109\discretionary{/}{%
}{/}TVCG\hspace{.1pt}\discretionary{.}{%
}{.}\hspace{.4pt}2007\hspace{.1pt}\discretionary{.}{%
}{.}\hspace{.4pt}70589}}}


\bibitem{williamson1992dynamic}
C.~Williamson and B.~Shneiderman.
\newblock {The Dynamic HomeFinder: Evaluating dynamic queries in a real-estate information exploration system}.
\newblock In {\em {Proceedings of the 15th annual international ACM SIGIR conference on Research and development in information retrieval}}, pp. 338--346, 1992. \href{https://doi.org/10.1145/133160.133216}
{doi: {{%
10\hspace{.1pt}\discretionary{.}{%
}{.}\hspace{.4pt}1145\discretionary{/}{%
}{/}133160\hspace{.1pt}\discretionary{.}{%
}{.}\hspace{.4pt}133216}}}


\end{thebibliography}








\end{document}